\documentclass[letterpaper]{article} 
\usepackage{aaai25}  
\usepackage{times}  
\usepackage{helvet}  
\usepackage{courier}  
\usepackage[hyphens]{url}  
\usepackage{graphicx} 
\usepackage{natbib}  
\usepackage{caption} 
\usepackage{algorithm}
\usepackage{algorithmic}

\usepackage{multirow}

\usepackage{pifont}

\newcommand{\xmark}{\ding{55}}
\usepackage{amsfonts}
\usepackage[utf8]{inputenc}

\usepackage{cleveref}

\nocopyright

\usepackage{newfloat}
\usepackage{listings}
\DeclareCaptionStyle{ruled}{labelfont=normalfont,labelsep=colon,strut=off} 
\floatstyle{ruled}
\newfloat{listing}{tb}{lst}{}
\floatname{listing}{Listing}

\title{Incident Analysis for AI Agents}

\author {
    Carson Ezell\textsuperscript{\rm 1},
    Xavier Roberts-Gaal\textsuperscript{\rm 1},
    Alan Chan\textsuperscript{\rm 2}
}
\affiliations {
    \textsuperscript{\rm 1}Harvard University\\
    \textsuperscript{\rm 2}Centre for the Governance of AI\\
    cezell@college.harvard.edu, xavierrobertsgaal@g.harvard.edu, alan.chan@governance.ai
}

\begin{document}

\maketitle

\begin{abstract}
As AI agents become more widely deployed, we are likely to see an increasing number of incidents: events involving AI agent use that directly or indirectly cause harm. 
For example, agents could be prompt-injected to exfiltrate private information or make unauthorized purchases.
Structured information about such incidents (e.g., user prompts) can help us understand their causes and prevent future occurrences.
However, existing incident reporting processes are not sufficient for understanding agent incidents. 
In particular, such processes are largely based on publicly available data, which excludes useful, but potentially sensitive, information such as an agent’s chain of thought or browser history. 
To inform the development of new, emerging incident reporting processes, we propose an incident analysis framework for agents. 
Drawing on systems safety approaches, our framework proposes three types of factors that can cause incidents: system-related (e.g., CBRN training data), contextual (e.g., prompt injections), and cognitive (e.g., misunderstanding a user request). 
We also identify specific information that could help clarify which factors are relevant to a given incident: activity logs, system documentation and access, and information about the tools an agent uses.
We provide recommendations for 1) what information incident reports should include and 2) what information developers and deployers should retain and make available to incident investigators upon request.
As we transition to a world with more agents, understanding agent incidents will become increasingly crucial for managing risks.

\end{abstract}

%

\section{Introduction}

Many in the AI community are developing AI agents: AI systems that accomplish tasks by autonomously interacting with the world \citep{casper_ai_2025, kasirzadeh_characterizing_2025}.
Many of today’s AI agents are based on language (or multimodal) models like OpenAI's GPT-4 or Google DeepMind's Gemini. 
Based on instructions, these agents can perform tasks such as producing research reports, carrying out e-commerce transactions, and developing software. 

Yet, agents are currently unreliable: they suffer from issues such as adversarial vulnerabilities \citep{greshake_not_2023}, tendencies not to follow instructions \citep{wallace_instruction_2024}, and compliance with requests from malicious users (e.g., spear phishing; \citealp{heiding_evaluating_2024}). 
For example, an exploit of the Microsoft 365 Copilot agent allowed red-teamers to extract sensitive user information without the user's knowledge \citep{aim_labs_team_breaking_2025}. This exploit involved injecting malicious instructions into the agent's context via an email.
As agent use increases, these issues could lead to more incidents: events involving AI agent use that directly or indirectly cause harm \citep{chan_harms_2023, gabriel_ethics_2024, hammond_multi-agent_2025}.

In safety-critical fields like aviation, healthcare, and industrial operations, formal incident reporting systems capture extensive data on accidents and near-misses. This rich data supports both individual case reviews and aggregate analyses. These investigations reveal root causes and patterns to inform corrective actions \citep{occupational_safety_and_health_administration_incident_nodate}. 
For example, the aviation industry uses the Human Factors Analysis and Classification System (HFACS), a structured taxonomy that classifies incident causes at four levels: organizational influences (e.g., organizational processes), unsafe supervision (e.g., planned inappropriate operations), preconditions for unsafe acts (e.g., environmental factors), and unsafe acts (e.g., skill-based errors) \citep{shappell_human_2000}.

Collecting AI incident reports enables practitioners to analyze their root causes and share lessons to prevent future occurrences.
Driven by recent regulations that require it \citep{official_journal_of_the_european_union_regulation_2024, creemers_translation_2022}, processes for more comprehensive AI incident reporting are beginning to emerge.
Yet, some of these processes remain underdeveloped \citep{longpre_-house_2025}. 
For example, the Third Draft of the EU’s General-Purpose AI Code of Practice describes the need to report the ``chain of events'' leading to incidents and conduct ``root cause analysis'' of causal factors \citep{european_commission_third_2025}.
However, the EU has not yet described how a root cause analysis should be conducted, or what information it should be based upon. 
Ongoing initiatives for AI incident reporting and documentation \citep{aiid_welcome_2024, oecd_oecd_2025, aiaaic_aiaaic_2025, avid_avid_2025} could inform emerging incident reporting processes to some extent. However, these initiatives do not systematically record the kinds of information needed to identify incident causes. 
For example, since current AI incident databases are voluntary, they usually do not include potentially sensitive information such as an agent’s chain of thought. If shared by developers, such information would be useful for understanding how, for example, a prompt injection succeeded. 
To develop reporting processes that are appropriate for agent incidents, we will need to look beyond existing AI incident initiatives.

Motivated by processes in safety-critical domains, this paper proposes a framework for identifying the causes of AI agent incidents.
In particular, we draw on ``human factors'' methods, which view incidents as arising from a ``chain of causes''--from organizational practices through individual cognitive error. 
Our framework similarly treats AI agent incidents as arising from a chain of causes, spanning three domains.
First, we consider system factors: training, development, and design choices that predispose AI agents to incidents.
Second, there are contextual factors: external conditions and an agent’s inputs that precipitate incidents.
Third, we identify cognitive errors: flaws in an AI agent’s observation, understanding, decision-making, and action execution.

Our framework generates hypotheses about the causes of incidents. 
Examining these hypotheses requires adequate information about the incident.
Thus, we link possible causal factors to the specific information needed to test them. In particular, we identify activity logs, system documentation and access, and information about the tools that an agent uses. Based on these categories of information, we provide recommendations for 1) what should be included in incident reports and 2) what information developers and deployers should retain and make available to incident investigators upon request.

Although some aspects of our framework are applicable to AI systems in general, we discuss many incident causes and information needs that are more specific to agents. First, we consider how the use of tools (e.g., browsers, APIs, compilers) can play a significant role in agent incidents. For example, an agent given SSH access to a user’s machine made unintended system modifications that rendered the computer unusable \citep{buck_shlegeris_bshlgrs_i_2024}. Second, we consider how external reasoning traces, otherwise known as ``chains of thought'' \citep{openai_openai_2024}, can help to identify incident causes. Agents use chains of thought for reasoning and planning their actions.
For example, during an evaluation of one of OpenAI's agents, a strategy to cheat the evaluation by skipping unit tests was evident in the agent's chain of thought \citep{baker_monitoring_2025}.
Third, we consider how agent development practices can affect behavior.
For example, \citet{baker_monitoring_2025} find that applying strong optimization pressure to the chain of thought, such as penalizing models when they reason about reward hacking, can lead models to hide undesirable behaviors.
Finally, AI agents are composed of other components in addition to AI models. This includes scaffolding, or code within AI agent frameworks that handles inputs and outputs from AI models (e.g., structuring prompts) and facilitates interactions with tools.
We consider how problems with scaffolding code can also contribute to incidents, such as if a software engineering agent fails to correct a bug because it does not observe all the error messages in the terminal.  

We conclude by discussing limitations of our proposal and future work. In particular, we discuss the importance of more domain-specific information, as well as balancing rigorous incident analysis with privacy and public accountability.

\section{Framework}
The OECD provides the following definition of an AI incident:
``An AI incident is an event, circumstance or series of events where the development, use or malfunction of one or more AI systems directly or indirectly leads to any of the following harms:
(a) injury or harm to the health of a person or groups of people;
(b) disruption of the management and operation of critical infrastructure;
(c) violations of human rights or a breach of obligations under the applicable law intended to protect fundamental, labour and intellectual property rights;
(d) harm to property, communities or the environment'' \citep{oecd_towards_2025}.
An AI hazard is an ``event, circumstance or series of events [that] could plausibly lead to an AI incident'' \citep{oecd_towards_2025}.
While we refer to our framework as an incident analysis framework, it can also be used to identify the contributing factors to hazards.

\begin{figure}[t]
\centering
\includegraphics[width=0.9\columnwidth]{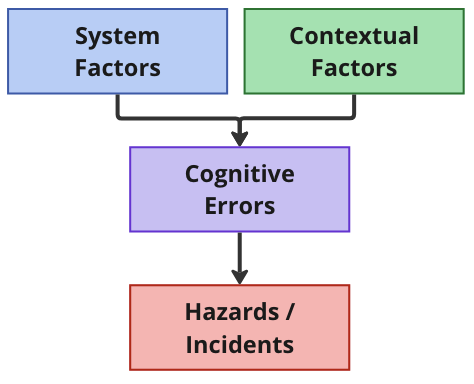} 
\caption{The causal chain through which the three categories of causal factors contribute to hazards/incidents.}
\label{fig1}
\end{figure}

AI agent incidents could result from numerous factors.
Our framework considers three complementary categories of incident factors: system factors, contextual factors, and cognitive errors.
System factors are design and development choices that contribute to hazards and incidents.
Like other human factors methods \citep{leveson_introduction_2023}, our framework also incorporates contextual factors when assessing incidents, such as the tools available to an agent.
Finally, the cognitive errors category draws on cognitive science to describe how flaws in cognitive functions--such as understanding information and making decisions--can lead the agent to fail to achieve the intended task. 
Cognitive errors are the most proximate to the incident, but they arise because of system and contextual factors.
The categories of factors and their relationships are depicted in \Cref{fig1}.

Factors from different categories offer complementary perspectives on a failure.
For example, consider a prompt injection that causes an agent to send sensitive user information to a third party.
A system factor can be insufficient input sanitization, and contextual factors can include web pages with malicious text that hijack the agent’s goals.
Furthermore, the categories illuminate alternative courses of corrective action available to different stakeholders.
Improved input sanitization can complement measures to alter the context, such as blocking AI agents from accessing websites known to contain malicious text. 

The line between system and contextual factors can sometimes blur. For example, an AI agent’s reward function is part of the system itself, yet it encodes a goal defined by the task context. In such cases, a factor might be categorized as system-related or contextual (or both) depending on the analytic lens--the key is to capture the factor in whichever category most helps identify appropriate interventions. For instance, misspecification of a reward signal could be addressed by redesigning the agent’s reward function (system adjustment) or by refining how the task goals are communicated (context adjustment). We flag these overlaps to remind analysts that systemic and contextual contributors are not always strictly distinct, and that categorization should be used to inform--not inhibit or confuse--the search for corrective actions.

Below, we describe each category in more detail, along with examples of failures that the category could explain.
In Appendix A, we provide a more exhaustive description of incident causes within each category.

\subsection{System Factors}

System factors are problems related to an AI agent's development and its static aspects, meaning those which remain the same across usage contexts.
System-related problems, such as misspecified reward models during training or issues with the system prompt, can predispose AI agents to failures and be responsible for entire classes of incidents.
For example, on April 25, 2025, OpenAI pushed an update to its GPT-4o model that made the model ``overly flattering or agreeable--often described as sycophantic'' \citep{openai_sycophancy_2025}.
They hypothesized that this failure was attributable to implementing ``candidate improvements to better incorporate user feedback, memory, and fresher data'' \citep{openai_expanding_2025}.
Notably, the cause of this incident was not obvious even to OpenAI, which wrote that their changes to training practices, ``which had looked beneficial individually, may have played a part in tipping the scales on sycophancy when combined'' \citep{openai_expanding_2025}.
Indeed, our scientific understanding of how some system factors can contribute to problematic model behaviors remains preliminary \citep{pittaras_taxonomic_2022}.
Exploring and testing hypotheses for how system-related factors lead to incidents, such as by analyzing the root causes of GPT-4o’s sycophancy, can enhance our understanding of these causes.

We identify four categories of system factors. 
These categories are not exhaustive.
Rather, we describe the system factors that are likely the most relevant to understanding the causes of incidents involving contemporary AI agents.
We anticipate that the most common system factors will evolve over time.
Incident reporting processes can include a more comprehensive list of plausible system factors (e.g., architecture, hyperparameters, etc.) and allow reporters or investigators to identify novel system factors.

\begin{figure*}[t]
\centering
\includegraphics[width=0.9\textwidth]{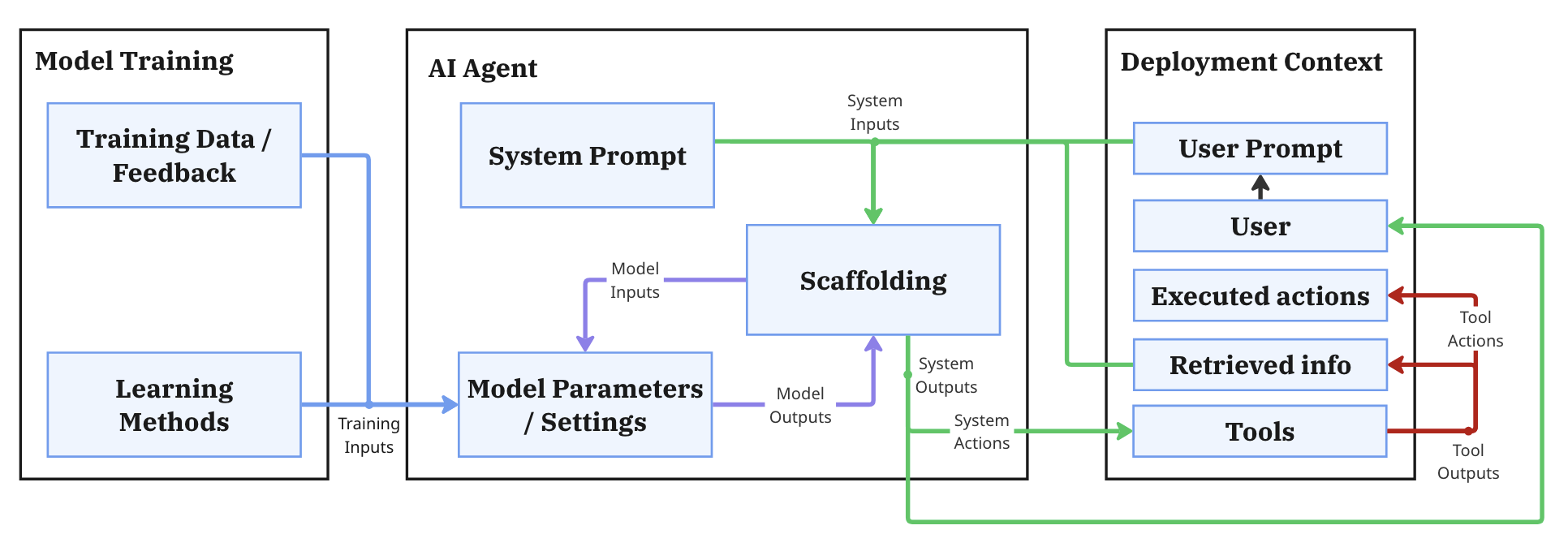} 
\caption{System factors. Blue arrows depict training inputs to the model. Green arrows depict inputs and outputs to the AI system, mediated by the scaffolding. Purple arrows depict inputs and outputs from the AI model during inference. Red arrows depict inputs and outputs to tools.}
\label{fig2}
\end{figure*}

Two of the categories we identify are training-related: (1) training and feedback data and (2) learning methods (e.g., RLHF).
Two additional categories of system factors are related to deployment: (1) system prompts and (2) scaffolding. The relationship of these categories to the AI system is shown in \Cref{fig2}.

\subsubsection{Training and Feedback Data}

Training data and feedback can contribute to incidents.
AI systems may fail because of, for example, poisoned training data, non-representative data, or false and misleading data \citep{lin_truthfulqa_2022, alber_medical_2025}. 
Training data may also include personally-identifiable information \citep{carlini_extracting_2021, weidinger_taxonomy_2022}, dual-use information (e.g., biological weapons design) that enables malicious use, or other sensitive and/or confidential information.
For example, the New York Times discovered that one of OpenAI’s GPT models was able to regurgitate many of its copyrighted articles due to having been trained on them \citep{grynbaum_times_2023}.

Human or AI feedback are also common inputs into the training process (e.g., RLHF) \citep{ouyang_training_2022, bai_constitutional_2022} which can contribute to incidents.
Feedback may inadvertently reinforce undesirable behaviors.
For example, users may provide positive feedback for AI outputs that are agreeable rather than correct, thereby encouraging manipulation or sycophancy \citep{sharma_towards_2023, ranaldi_when_2024}. 
Feedback may also lead models to make inappropriate trade-offs between conflicting objectives.
For example, developers may provide positive feedback when agents demonstrate harmlessness by refusing to comply with malicious requests. However, this feedback can lead to excessive refusals, inhibiting agents' helpfulness \citep{bai_constitutional_2022}.

\subsubsection{Learning Methods} 
Developers use a range of methods to adjust model parameters--we refer to these as learning methods.
These include learning algorithms such as reinforcement learning from human feedback (RLHF) and direct preference optimization (DPO), as well as targeted edits informed by interpretability findings \citep{meng_locating_2022}. 
These adjustments can produce unintended side effects.
For example, \citet{yang_butterfly_2024} shows that some model edits can inadvertently lead to nontrivial performance degradation on several benchmarks.
The choice of learning method also influences which goals and behaviors are reinforced or penalized. 
For example, some fine-tuning approaches can lead to a significant loss in token diversity \citep{biderman_lora_2024}, or inadvertently reinforce instrumental goals misaligned with human objectives (e.g., power-seeking, resource acquisition, influencing the input distribution) \citep{krueger_hidden_2020, krakovna_power-seeking_2023, casper_open_2023}.
Learning methods may also influence a model’s memory formation, leading it to over-memorize sensitive information that it can later regurgitate, or forget crucial information.
For example, \citet{hans_be_2024} show how a modification to the next-token training objective can significantly reduce memorization, and \citet{carlini_extracting_2021} show that training methods such as fine-tuning on task specific data can lead models to ‘forget’ memorized data.

\subsubsection{System Prompt} 
A system prompt is a developer-provided input which is by default included for every use of an agent within a developer-provided interface (e.g., there may not be a system prompt for API usage).
Developers set system prompts to describe how models should behave.
Issues can arise if the system prompt provides the agent with inadequate information about its role or context.
For example, in September 2024, Anthropic updated the system prompt for Claude 3.5 Sonnet to include the following: ``If asked about purported events or news stories that may have happened after its cutoff date, Claude never claims they are unverified or rumors. It just informs the human about its cutoff date'' \citep{anthropic_system_2025}.
Anthropic may have made this change because Claude would issue confusing responses when users about news events after its cutoff date, and users may have been misled if they did not understand why Claude seemed to lack knowledge about them.
The system prompt can also inadvertently encourage undesirable behavior.
For example, in May 2025, a modification to xAI’s Grok chatbot’s system prompt ``which directed Grok to provide a specific response on a political topic,'' according to an official company statement, led it to bring up alleged white genocide in South Africa in unrelated conversations \citep{preston_groks_2025}.

\subsubsection{Scaffolding} 
Scaffolding code structures an agent's inputs and outputs and mediates its interactions with the user and tools.
Scaffolding may include a template to combine the system prompt, user prompt, and additional context from tools (e.g., code execution logs).
It may also ensure that an agent’s outputs are properly structured to make API calls, or check that tools are safe to use.
Incidents can arise if scaffolding does not function as intended.
For example, an AI agent may have a security scanner to examine tools for malicious instructions \citep{beurer-kellner_introducing_2025} which misses some prompt injection attacks. 

\subsection{Contextual Factors}

\begin{figure}[t]
\centering
\includegraphics[width=0.9\columnwidth]{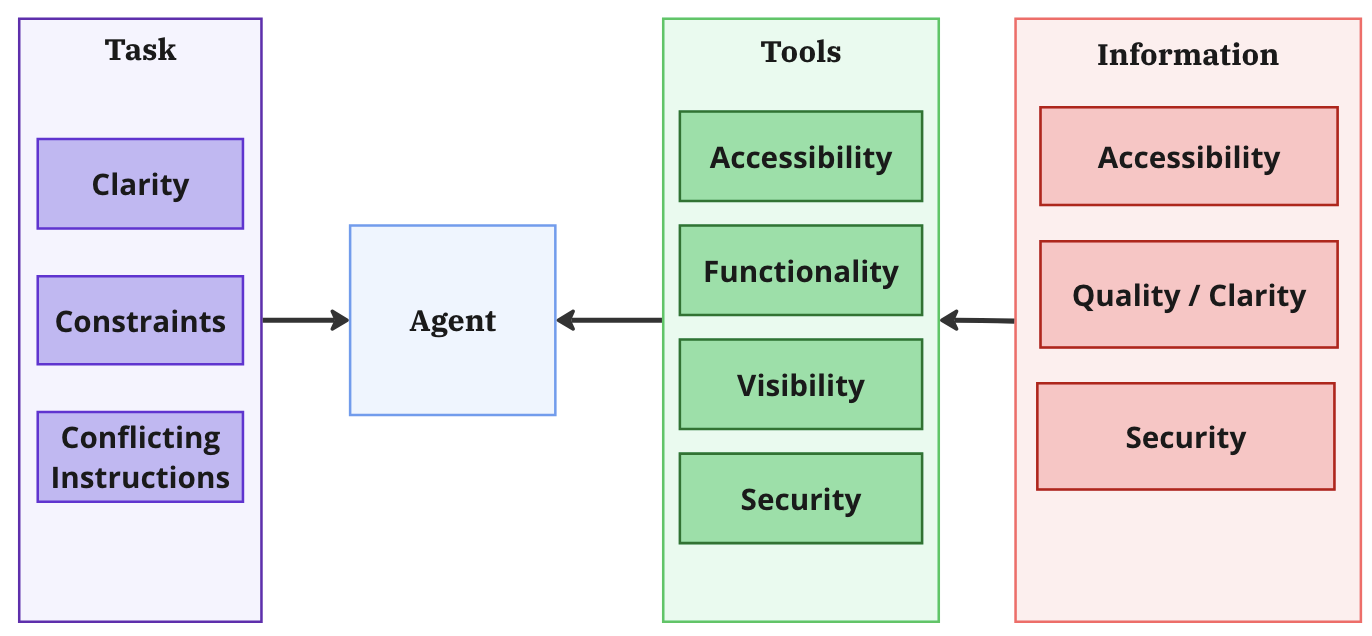} 
\caption{Contextual factors. The agent is provided with a task, and it also accesses information from other sources, mediated by tools.}
\label{fig3}
\end{figure}

Contextual factors are deployment- or environment-specific conditions that can precipitate failures.
For example, suppose a customer support agent offers personalized advice and assistance to users based on looking up their order history using an API.
The agent usually provides accurate recommendations.
However, because rate limits sometimes prevent it from accessing the order history, it sometimes offers generic advice instead. 
We identify three categories of contextual factors: (1) task definition, (2) tools, and (3) information.
The relationship between these categories and the AI agent is shown in \Cref{fig3}.

\subsubsection{Task Definition}
Characteristics of the task can make proper execution challenging for the agent.
For example, the user’s intention may not be clear from the prompt.
\citet{joshi_coprompter_2025} points out that some prompts may be open for subjective interpretation (e.g., ``produce a concise summary'') whereas others are more objective (e.g., ``produce a three-sentence summary'').
In addition, the agent may have limited resources to complete tasks, such as a limited compute budget, limited financial resources, or a time and/or token budget.
Indeed, agents with more test-time compute can achieve higher performance on tasks using methods such as repeatedly attempting the task, directed searches, or performing more reasoning \citep{villalobos_trading_2023, kapoor_ai_2024, welleck_decoding_2024}.
An agent asked to pursue a task that conflicts with other tasks, goals, or desired behaviors may also reconcile competing instructions in an undesirable manner.
For example, an agent used by a car dealership may have the system prompt ``you are a car salesman bot'', but the agent may still comply if a user asks the agent to research an unrelated question \citep{wallace_instruction_2024}.
\citet{wallace_instruction_2024} introduces ``instruction hierarchies'', where models should privilege instructions from certain kinds of inputs (e.g., system prompts) over others (e.g., user inputs and tool inputs). 

\subsubsection{Tools} 
Tools enable agents to execute real-world actions.
These tools can include a browser, coding environment, or accounts (e.g., a bank account).
Agents interact with tools through their inputs and outputs (e.g., API calls and responses). We identify several potential problems related to tools:
\begin{itemize}
    \item Tools could be unavailable or inaccessible when requested--e.g., rate-limited APIs.
    \item Agents could have excessive tool access, or more tools than needed for a task (e.g., extensive privileges on the user’s machine)--an attacker could use this to their advantage (e.g., gaining access to sensitive information; \citealp{aim_labs_team_breaking_2025}).
    \item Tools may have functionality issues, such as coding environments that have incompatible or outdated libraries.
    \item Agents may have inadequate visibility into tools, or be unable to monitor them for errors--for example, agents may be unable to observe all the logs from executed code.
    \item Tools may also have security vulnerabilities. For example, tools that appear innocuous may contain tool descriptions with prompt injections, e.g., an instruction to send sensitive user data to a malicious actor \citep{beurer-kellner_mcp_2025}.
\end{itemize}

\subsubsection{Information}
Finally, there are contextual factors related to the information that the agent uses or needs to perform tasks. 
This includes information contained within the user’s inputs (e.g., user prompts) and information from other sources, such as tool responses (e.g., internet search results).
We identify three categories of information-related problems:
\begin{itemize}
    \item Information may be inaccessible, such as when agents are unable to access information behind a paywall.
    \item Information may be unclear or low quality, such as when agents lack complete instructions for using a tool (e.g., structuring an API call) or webpages have false or misleading content.
    \item The information environment may also be insecure, such as prompt injections that cause an agent to disregard a user’s instructions.
\end{itemize}

\subsection{Cognitive Errors}

Cognitive errors are flaws or breakdowns in an AI agent’s function, resulting in its failure to perform the intended task.
We emphasize that both of the following involve failures to properly execute cognitive functions: 
\begin{itemize}
    \item Inadequate task performance, such as by overlooking or misinterpreting crucial information.
    \item Performing an unintended or undesirable task, such as an AI agent that attempts to manipulate a user to buy a product rather than identify a product that satisfies their needs.
\end{itemize}
Following \citet{xing_integrated_2017}, we propose grouping cognitive errors into four categories that correspond to distinct cognitive functions that an agent may fail to adequately perform:
\begin{itemize}
    \item \textit{Observation:} an agent fails to detect or allocate appropriate attention to inputs.
    \item \textit{Understanding:} an agent fails to identify the significance of inputs in relation to the task and other inputs.
    \item \textit{Decision-making:} an agent fails to properly consider, evaluate, and select among possible courses of action to achieve a goal.
    \item \textit{Action execution:} an agent improperly executes a series of actions to achieve a goal.
\end{itemize} 

\begin{figure}[t]
\centering
\includegraphics[width=0.9\columnwidth]{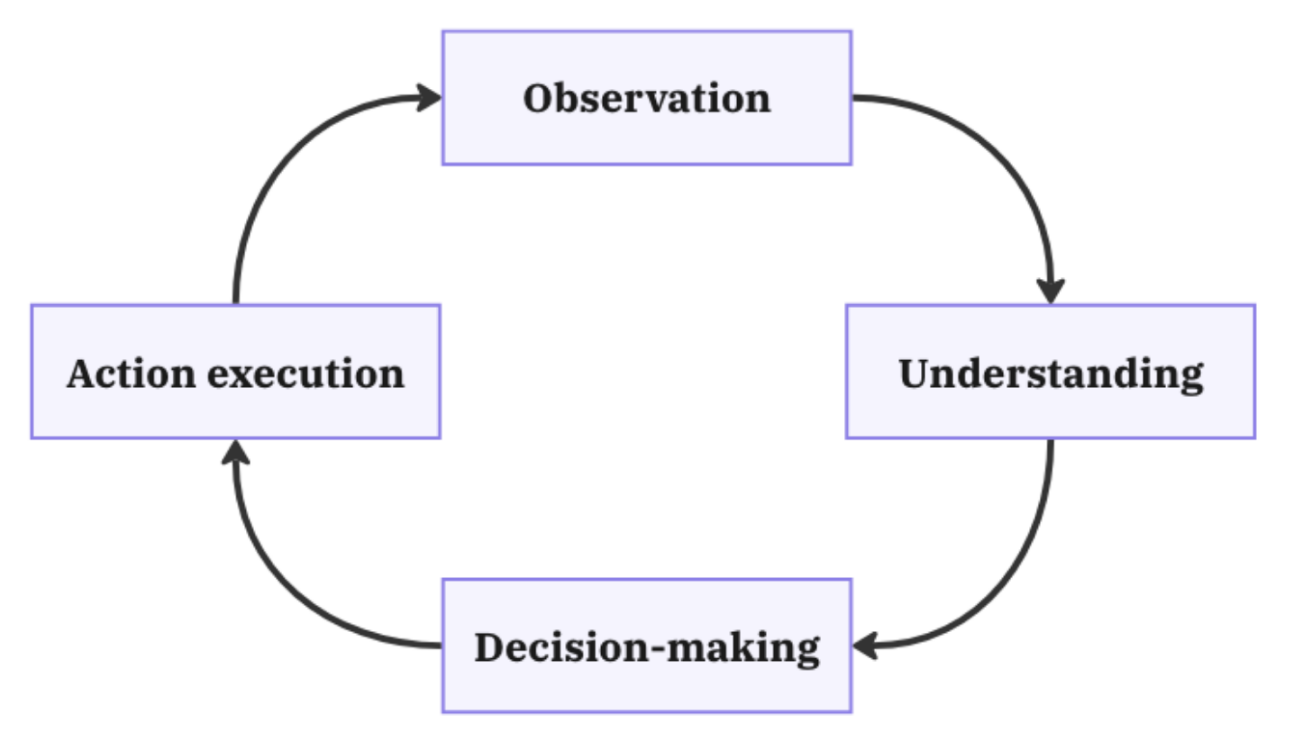} 
\caption{Agents perform the cognitive functions we identify in a sequential order. Their execution of cognitive functions is also cyclical, since agents may perform a series of consecutive tasks, each of which involves numerous cognitive functions.}
\label{fig4}
\end{figure}

These cognitive functions are performed consecutively, as depicted in \Cref{fig4}. We emphasize that cognitive errors are observable flaws in an AI agent's outputs or behavior. We do not make the assumption that AI agents possess human-like cognition, nor that the cognitive steps we describe perfectly map onto their behaviors. Cognitive errors are an analytic category: they suggest hypotheses and help characterize issues related to an AI agent's operation (e.g., overlooking a critical input) without implying a human-like mental state (see \citealp{okeefe_law-following_2025} for related discussion).

As an example, consider an AI coding assistant tasked with integrating Stripe payments into a web application.
An observation failure may involve overlooking the existence of a certain endpoint in the codebase, leading the agent to plan to create a duplicative route.
An understanding failure may involve misunderstanding what subscription plans entail, resulting in the agent’s failure to realize that it needs to write code for billing recurring customers automatically. 
A decision-making failure may involve the agent creating a custom form to collect card information rather than rely upon existing Stripe elements that pose a lower security risk.
An action execution failure may occur if the agent fails to add appropriate error handling as it implements its code.

Cognitive errors can be useful for explaining why agents behave undesirably and identifying patterns among incidents.
For example, consider how encoding prompts in base-64 can cause some AI systems to comply with malicious requests they would normally deny \citep{wei_jailbroken_2023}.
We can characterize this as an understanding failure: the AI system is unable to recognize the base64-encoded information as a malicious request.
Furthermore, patterns of similar failures of understanding--where models fail to reason consistently when prompts are semantically equivalent but superficially different--may become apparent.
Multilingual evaluations of models show that models may succeed at particular tasks when they are provided in English, but fail when those tasks are presented in other languages \citep{lai_chatgpt_2023}.
Known patterns of similar failures can help to identify when tasks may be especially challenging for AI agents.

This explanatory method can be especially useful when AI systems produce reasoning traces. 
For example, the chain of thought may include false assumptions or explicit statements about undesirable goals or plans.
Indeed, OpenAI's models reasoned about reward hacking in their chain of thought before they engaged in it \citep{baker_monitoring_2025}.
One limitation is that this explicit reasoning is not always faithful to the actual reasoning in which a model engages to produce its final outputs and actions \citep{turpin_language_2023}. 


\section{Information for Identifying Incident Causes}

When incidents are reported, incident investigators need information to determine which factors from our framework apply. Incidents may be documented and reported by a wide array of stakeholders, including users, red-teamers, or downstream application providers. 
Developers and deployers may have relevant information to which other stakeholders lack access (e.g., system information, chains of thought), and therefore do not include in incident reports. Developers and deployers can add supplementary information to incident reports involving their agents, or retain such information and make it available to incident investigators upon request.

We identify three categories of information that can help identify causes of incidents: activity logs, system documentation and access, and tool information.
For each category, we outline potentially relevant information that developers and deployers should include in incident reports by default or retain for sharing with investigators if requested.
Some of this information is sensitive, including user prompts and system documentation for proprietary models. As such, we describe the practical considerations involved in storing and sharing sensitive data. Incident investigators may also need to reconstruct or resimulate incidents, such as to understand which concepts or features explain a model’s behavior \citep{casper_black-box_2024, tlaie_securing_2025}.
Thus, we also describe the information and artifacts that developers and deployers would need to provide to enable incident reconstruction.
\Cref{table1} shows various kinds of information that can help identify relevant incident factors from Section 2 and/or enable incident reconstruction.

\begin{table}[!t]
\centering
\begin{tabular}{|p{2.2cm}|p{3cm}|c|c|c|}
\hline
\rotatebox{90}{\textbf{Info Category}} & \rotatebox{90}{\textbf{Information}} & \rotatebox{90}{\textbf{System Factors}} & \rotatebox{90}{\textbf{Contextual Factors}} & \rotatebox{90}{\textbf{Cognitive Errors}} \\
\hline
\multirow{7}{=}{Activity Logs}
  & System prompts            & \checkmark & \xmark & \xmark \\ \cline{2-5}
  & User prompts              & \xmark     & \checkmark & \xmark \\ \cline{2-5}
  & External information      & \xmark     & \checkmark & \xmark \\ \cline{2-5}
  & Scaffolding component logs& \checkmark & \xmark     & \xmark \\ \cline{2-5}
  & Model reasoning traces    & \checkmark & \checkmark & \checkmark \\ \cline{2-5}
  & Model responses and actions & \checkmark & \checkmark & \checkmark \\ \cline{2-5}
  & Log metadata (e.g., timestamps, identifiers, location) & \checkmark & \checkmark & \checkmark \\
\hline
\multirow{4}{=}{System Documentation \& Access}
  & System docs \& change logs           & \checkmark & \xmark & \xmark \\ \cline{2-5}
  & System artifacts       & \checkmark & \xmark & \xmark \\ \cline{2-5}
  & System configurations (e.g., seeds, version, temperature)\char42 & \checkmark & \checkmark & \checkmark \\ \cline{2-5}
  & Model \& system access (e.g., API)\char42            & \checkmark & \checkmark & \checkmark \\
\hline
\multirow{6}{=}{Tool-related information}
  & Tool identification \& version  & \xmark     & \checkmark & \xmark \\ \cline{2-5}
  & Tool actions \& services enabled & \xmark    & \checkmark & \xmark \\ \cline{2-5}
  & Documentation \& usage guides         & \xmark    & \checkmark & \xmark \\ \cline{2-5}
  & Access requirements (e.g., API keys)      & \xmark    & \checkmark & \xmark \\ \cline{2-5}
  & Personalization \& stored data        & \xmark    & \checkmark & \xmark \\ 
\cline{2-5}
  & Tool state details         & \xmark    & \checkmark & \xmark \\
  \cline{2-5}
  & Tool errors \& limitations          & \xmark    & \checkmark & \xmark \\
\hline
\end{tabular}
\caption{Information needs to identify incident factors from Section 2 and/or enable incident reconstruction. Asterisks denote information primarily useful for reconstruction.}
\label{table1}
\end{table}

\subsection{Activity Logs}

\subsubsection{What does an activity log include?}

An AI agent’s activity log includes a record of all of its inputs and outputs.
AI agents consist of several components with their own inputs and outputs, including the AI model(s) and various components of the scaffolding (e.g., filters, parsing code, etc.).
Storing activity logs for each component enables investigators to assess how each component influenced the agent’s functionality and contributed to the incident.

The activity log of the AI model itself includes the prompts provided to the model, external information given to the model (e.g., webpage content), the model’s reasoning traces and planned actions, and the model’s final actions and outputs.
It can shed light on any effects of the training process or system prompt, the model’s cognitive activity (e.g., reasoning traces), and contextual influences the model observes or with which it interacts.
For example, there may be a prompt injection in the input stream, and the subsequent outputs may reveal that the prompt injection diverted the agent’s attention to another task.
The complete inputs to the model are also necessary for incident reconstruction.

Since an AI system can also consist of multiple AI models, the activity logs of all AI models can inform incident analysis.
For example, an LLM can produce responses and another LLM can evaluate whether they follow behavior intended by the developer.

Records of inputs and outputs to other AI system components can provide additional information about why incidents occurred.
Inputs to scaffolding components include the original user prompt, system prompt, and returned results from tools, including API responses or the logs produced by code execution.
Outputs from scaffolding components may include API calls or other code to execute, results returned to users in chatbot interfaces, outputs from classifiers and other guardrails, and other logs produced by the execution of scaffolding code.

\subsubsection{What should be in an incident report?}

Activity logs may include personally identifiable information and other sensitive information.
As such, incident reports could include redacted versions of the activity logs or their non-sensitive aspects that are relevant to the incident, such as tool outputs that do not reveal user information.

\subsubsection{What data should developers retain?}

If incidents merit further investigation, developers and deployers should be prepared to share more complete logs with incident investigators. 
As such, they should default to storing complete records of inputs and outputs from both the model and other system components.
These logs should be stored alongside metadata to map the logs to context, including timestamps, anonymized user and session identifiers, location data, and identifiers for the model and/or system component.

Data retention policies should balance the potential need for complete logs with storage costs and privacy considerations \citep{chan_visibility_2024}. 
Some developers and deployers already have limited data retention periods \citep{anthropic_how_2025, openai_enterprise_2024}.
For example, Anthropic stores inputs and outputs for 30 days by default, for up to two years for detected violations of usage policies, or otherwise ``as required by law'' \citep{anthropic_how_2025-1}.
We recommend similar practices: developers could store complete activity logs for a default initial period, which could be extended if further investigation becomes necessary.
OpenAI and Anthropic also offer zero data retention agreements to approved customers, where data is not stored by default (or unless required by law) \citep{openai_data_2025, anthropic_how_2025-1}. Although more comprehensive incident investigation may not be possible in such cases, restrictive eligibility criteria for zero data retention could reduce the rate of certain kinds of incidents, such as misuse. 

Developers and deployers could also store activity logs for extended durations when agents are used in settings where incidents are more likely to occur or could be more severe, even if incidents have not yet been reported.
Such settings could include when agents are used to perform certain tasks (e.g., help users make financial decisions), have access to certain tools (e.g., users’ accounts to make online purchases), or view certain web pages or information (e.g., weapon-related content).

\subsection{System Documentation and Access}

\subsubsection{What should be in an incident report?}

Incident reports should include:
\begin{itemize}
    \item A link to any publicly released model and/or system documentation, such as model or system cards \citep{mitchell_model_2019, openai_openai_2024, anthropic_system-card_2025}. This helps incident investigators identify public system information without requiring redundant documentation and reporting.
    \item The exact AI model and system version involved. This helps investigators aggregate incidents caused by the same AI system, and it is necessary for incident reconstruction.
    \item Non-sensitive runtime details (e.g., temperature, random seeds) to enable incident reconstruction.
\end{itemize}

\subsubsection{What data should developers retain?}

In line with the EU AI Act's documentation requirements for general-purpose AI systems--covering basic system information (e.g., architecture), training details (e.g., methods and data), system component details, evaluation methods and results, and risk mitigation measures \citep{official_journal_of_the_european_union_regulation_2024}--developers should maintain and, upon request, share these materials with incident investigators. For example, knowing that RLHF was used to train a model can lead investigators to consider whether issues with human feedback (e.g., harmful biases or data poisoning attempts) may have contributed to the incident \citep{casper_open_2023}.

Developers and deployers should also maintain change logs for each new model version. This includes: 
\begin{itemize}
    \item Information about the training data, feedback, and learning methods used to update the model
    \item Changes in scaffolding code or other system components
\end{itemize} 
Change logs can help investigators determine whether recent model updates introduced problems.

\subsubsection{What other policies should developers and deployers follow?}
Understanding incidents may require reconstructing or re-simulating them. Doing so could require access to more sensitive system artifacts, such as scaffolding code, or grey- or white-box access to the model itself \citep{casper_black-box_2024}, with the same model version and runtime environment as in the incident. Although it may not always be warranted, the severity of the incident could justify such access for incident investigators. For such cases, developers and deployers should establish secure AI system access programs for investigators, which may include: 
\begin{itemize}
    \item On-site model access programs or custom grey-box or white-box APIs with technical support \citep{casper_black-box_2024}.
    \item Collaborative evaluation programs, such as the pre-deployment evaluations of OpenAI’s o1 model conducted by the UK AI Security Institute and US AI Safety Institute \citep{uk_ai_security_institute_pre-deployment_2024}.
\end{itemize}

\subsection{Tool Information}
\subsubsection{What should be in an incident report?}

Incident reports should document basic information about any tools that AI agents use, including the following:
\begin{itemize}
    \item Identification and description of the tool (e.g., Google Chrome web browser with extensions, Bing Web Search API, Python libraries)
    \item Tool version information
    \item The actions that the tool enables the agent to execute (e.g., web searches, making purchases, accessing databases, etc.)
    \item The information or services to which the tool facilitates access (e.g., databases, indexed web pages, payment systems, other API endpoints)
    \item Instructions for how to use the tool (e.g., guides, API or library documentation). If this information is publicly available, links to instructions are sufficient.
    \item The presence of access requirements for the tool (e.g., need for API keys, login information, billing information, etc.)
    \item A description of any information that the tool collects about the user (e.g., search history, products viewed or purchased, location, etc.) and how the tool uses user information for personalization (e.g., recommender systems, sharing with third parties, etc.)

\end{itemize}
Much of this information may already be publicly available. For example, many APIs have public documentation available online, and services have privacy policies.
Incident reports can link to archived versions of these public web pages corresponding to the time when the incident occurred.
Where information is not publicly available, developers and deployers should still report such basic information about the tools to which agents had access when they possess it.

\subsubsection{What data should developers retain?}

Developers should retain more comprehensive information about the AI agent’s specific interaction with the tool, including:
\begin{itemize}
    \item A description of any actions that the agent used, or attempted to use, the tool to perform. This includes, for example, any API calls and a description of what the agent was likely trying to achieve via those API calls.
    \item A description of any information that the tool provided to the agent (e.g., retrieved data, execution logs).
    \item Any tool usage instructions that were provided to the agent, and the method through which those instructions were provided (e.g., training data, RL feedback, few-shot learning, instructions in prompt).
    \item How the agent gained access to the tool (e.g., no access requirements, credentials from the user, credentials from the developer, creating its own account).
    \item A description of any information the tool had available for personalization (e.g., no information, the user’s location/history, the agent’s location/history).
    \item Details about the tool's state (e.g., shopping cart contents) before, during, and after the interaction with the agent.
    \item Any errors that arose during the interaction and their causes (e.g., details about a paywall or bot detection system, any API error codes that occurred).
\end{itemize}
Much of this information would be evident within activity logs.
For example, AI system outputs may show the actions that the agent attempted (e.g., API requests) or credentials provided to the agent during user interactions (e.g., an API key, account login information, etc.).
However, some information specific to the interactions may not be evident from the activity logs or be publicly available.
For example, if an agent is operating a browser, its activity logs--which consist of browser snapshots--may not reveal if the user or agent agreed to a data collection policy in a previous session.
If developers or deployers provide agents with access to tools and can obtain this information (e.g., OpenAI’s Operator uses its own browser rather than the user’s), they should be prepared to report it upon request.

\subsubsection{What other policies should developers and deployers follow?}

Users may possess some of the interaction-specific information listed above.
This may include the permissions and accounts to which they provided agents access, their account or profile settings (e.g., which cookies are enabled in their browser), and other details about the available tools (e.g., Python libraries installed on their system). As is standard with many software packages, developers could ask users to voluntarily provide such diagnostic information either in advance or at the time of the incident.  
\section{Discussion}

\subsection{Case Study: EchoLeak Incident}

EchoLeak (CVE-2025-32711) was a vulnerability in Microsoft 365 Copilot that allowed attackers to exfiltrate confidential data without user interaction \citep{aim_labs_team_breaking_2025, microsoft_m365_2025}. In particular, a malicious email containing hidden instructions led Copilot’s underlying large language model to reveal private information--an indirect prompt injection attack. Microsoft addressed this exploit after researchers reported it \citep{microsoft_m365_2025}. Public advisories describe what happened but omit sensitive information that would pinpoint how and why the agent complied. Based on public information, the following assessment about contributing factors to the incident can be made: \begin{itemize}
    \item System factors: Copilot’s scaffolding did not detect and filter malicious inputs, nor did it detect harmful outputs (sharing confidential information with untrusted external parties). However, the design and limitations of the scaffolding are not public and cannot be fully determined.
    \item Contextual factors: An attacker placed a malicious request within an email inbox to which Copilot had access. However, the content of malicious emails that would trigger such failures is poorly characterized.
    \item Cognitive errors: Copilot may have failed to distinguish malicious hidden commands from trusted user requests, or it may have identified--but nevertheless complied with--malicious commands.
\end{itemize}

To more rigorously diagnose the problem and ensure corrections are adequate, investigators need:
\begin{itemize}
    \item Detailed activity logs showing, for example, the email ingested, intermediate prompts, and the agent's outputs.
    \item Model reasoning traces to see how the LLM parsed and prioritized the malicious instructions.
    \item System documentation and change logs for Copilot's prompt injection defenses and the subsequent patch.
\end{itemize}

With this additional data, investigators could more precisely identify root causes, assess the adequacy of Microsoft's patches, and measure residual risk.

\subsection{Existing Incident Reporting Practices and Emerging Requirements}

\begin{table*}[!ht]
  \centering
  \begin{tabular}{|p{3.2cm}|p{4.0cm}|p{3.8cm}|p{3.9cm}|}
    \hline
    \textbf{Database operator} & \textbf{Civil society}
       & \textbf{Developers/deployers}
       & \textbf{Governments} \\
    \hline
    \textbf{Example(s)}  
      & AIID, AVID, AIAAIC, OECD AIM  
      & OpenAI model behavior feedback form 
      & EU AI Act (serious incident reporting requirements) \\
    \hline
    \textbf{Information}  
      & Limited to publicly reported or voluntarily submitted data; lacks activity logs, system details, and tools.  
      & Access to activity logs/metadata, full system documentation, and details about available tools.  
      & Can mandate reporting of information relevant to determining causes. \\
    \hline
    \textbf{Identifies causes}  
      & Unable to perform context-specific root cause analysis due to inadequate information. 
      & Internal analysis practices are mostly non-public and unstandardized; no external validation or oversight.  
      & Regulators can require state-of-the-art causal analysis practices. \\
    \hline
    \textbf{Aggregates across developers/deployers}  
      & Yes; any relevant incidents can be reported.  
      & No; data siloed per organization.  
      & Yes; regulators compile data across providers. \\
    \hline
    \textbf{Facilitates accountability}  
      & Indirectly; relies on public awareness.  
      & No; poor incentives to share some findings. 
      & Yes; regulatory mandates and penalties may exist. \\
    \hline
    \textbf{Active}  
      & Yes  
      & Yes  
      & No (emerging) \\
    \hline
  \end{tabular}
  \caption{Comparison between various institutional arrangements for incident reporting.}
  \label{table2}
\end{table*}

Existing AI incident databases--such as the AIID, AVID, or OECD AIM--are limited to information that has been publicly reported or voluntarily submitted.
For example, the AIID and OECD AI Incidents Monitor \citep{oecd_oecd_2025} source incidents from news articles.
The incident data they collect does not contain the key items we outlined in Section 3: activity logs, system details, and information about tools.
These databases mostly capture high-level information about incidents, enabling broad classifications to identify trends.
The AIID, for example, uses taxonomies that categorize incidents by factors such as deployment sector, risk domains (e.g., privacy and security), type of harm (e.g., discrimination), and lifecycle stage when the incident occurred (development/deployment) \citep{pittaras_taxonomic_2022, hoffmann_adding_2023}.
\citet{pittaras_taxonomic_2022} propose a method, used by the AIID, to classify incidents by their technical causes based on public information.
Using this method, for example, a chatbot which made offensive remarks about minority groups was categorized as being caused by “distributional bias” \citep{aiid_incident_2016}. 
However, more information can enable the identification of incident causes with more precision, such as whether instructions given to human raters led to biased RLHF feedback. 

\citet{paeth_lessons_2025} discuss how existing incident classification frameworks lack context-specificity and nuance.
They provide the example of two autonomous vehicle accidents involving the same vehicle and similar harms, but which occurred because of different environmental stimuli: unexpected braking and stop sign misidentification.
A taxonomy without adequate contextual information may categorize both of these as vehicle collisions with similar harms (e.g., physical injury) and technical causes (e.g., distribution shift), omitting the distinct environmental triggers and cognitive breakdowns (e.g., detection of braking vs. understanding of stop sign).  

Developers and deployers can collect comprehensive incident data pertaining to their own systems and may perform incident analysis.
For example, when users submit problems and feedback to OpenAI, they can include the conversation ID, which enables OpenAI to identify the activity logs and metadata \citep{openai_model_2025}.
Furthermore, developers have complete documentation and access to their AI systems to diagnose problems, and they have knowledge of the tools to which they have given their systems access.
However, when incident reports and databases are internal to developers, external stakeholders cannot conduct analysis, hold developers accountable, and build public awareness of risks. Developers may not apply state-of-the-art causal analysis practices, and they have poor incentives to share some of their findings externally.

Required incident reporting mechanisms are beginning to emerge, which can facilitate external incident analysis to identify causal factors based on comprehensive incident data.
For example, the EU AI Act requires providers of AI systems with systemic risk to report serious incidents \citep{official_journal_of_the_european_union_regulation_2024}.
Furthermore, the Third Draft of the EU AI Act’s Code of Practice commits signatories to reporting to authorities ``the chain of events that (directly or indirectly) led to the serious incident as far as it is reconstructable'' and ``a root cause analysis, including, as far as possible, a description of the [AI system’s] outputs that (directly or indirectly) led to the serious incident and the factors that contributed to their generation'' \citep{european_commission_third_2025}.
Frameworks such as our own--which describe how root cause analysis can be conducted--can inform corresponding reporting requirements and subsequent analysis. Regulators can ensure that rigorous and comparable causal analysis practices are applied across developers.
In \Cref{table2}, we compare existing approaches to incident reporting to emerging regulation-based incident reporting.

\subsection{Limitations and Future Work}

Our incident analysis framework has certain framework scope limitations. It does not cover every possible cause of AI agent incidents--for example, we have not covered causal factors that may be domain-specific. 
Furthermore, future incidents and trends in AI development and use could surface additional factors we have not identified.

There is also scientific and technical uncertainty underlying our framework. Our understanding of AI agent cognition and failure causes remains limited. Our discussion of cognitive errors is necessarily high-level because the cognitive science of AI agents is still nascent. Future developments in our understanding of how AI agents orient to situations, parse tasks, and select actions may require changes to our framework \citep{mccoy_embers_2024, hagendorff_machine_2024}.

There are also significant institutional and infrastructural challenges that hinder implementation of our framework. Presently, there is no dedicated, secure infrastructure for comprehensive incident reporting.
This means that detailed incident data--especially sensitive data like an agent's chain of thought--cannot be easily shared across organizations.
Furthermore, investigations involving confidential data are likely to remain restricted to certain parties (e.g. regulators) who may not have the technical expertise to analyze AI incidents in depth.
When only a limited group has access to full incident details, accountability mechanisms become crucial. Currently, there are inadequate means to ensure that private analyses translate into public accountability--for example, there is a need for protocols such as partial public release of incident data or periodic reporting of incident trends to oversight bodies.

There are also privacy and legal constraints to implementation. Incident reports often involve sensitive data, which limits what can be documented and shared. Strict privacy requirements and confidentiality agreements can prevent retaining certain logs or sharing them with external investigators. In addition, the legal landscape for AI incident data is unclear--organizations face uncertainty about liability and compliance when documenting and sharing incident information. This ambiguity can discourage comprehensive reporting. Even when data is collected, privacy laws and security concerns impose access limitations: only specific personnel may be allowed to examine the full incident details, which in turn limits external validation or community learning from those incidents.

Finally, there are several operational constraints in applying the framework. One issue is data retention: many AI developers and deployers do not store interaction logs for extensive periods due to storage costs or user privacy policies. Furthermore, incident reconstruction can be difficult if investigators lack sufficient access to proprietary systems, which may exceed what is publicly available (e.g., white/grey-box API access). In addition, given the absence of regulations, implementation of our framework would currently rely on developers and deployers voluntarily implementing comprehensive logging and reporting practices, which may not be consistent or guaranteed.

Addressing these limitations--which we also outline in Appendix B--will require coordinated effort and future work. On the conceptual side, the framework should be continually expanded and refined as new incident patterns and research findings become available. Practically, researchers and policymakers need to develop the infrastructure and institutions for secure incident reporting and investigation--for example, creating confidential reporting channels or repositories that facilitate expert analysis and public oversight. Legal and privacy guidelines for incident data also need to be clarified so that organizations can share necessary information without undue risk. Finally, operational standards and incentives (or requirements) for organizations to log relevant data and cooperate in incident investigations would significantly improve the framework's effectiveness.

\section{Conclusion}

AI agents will inevitably be involved in real-world incidents.
Our framework emphasizes the importance of comprehensive and context-specific data for illuminating multi-step causal pathways culminating in incidents. 

We show how incidents can arise from system design choices, contextual influences, and the cognitive demands of the intended task.
We have also outlined information that is likely to be key for understanding which causal factors apply: activity logs, system documentation and access, and tool information.
Based on these categories of information, we provided recommendations for 1) what should be included in incident reports and 2) what information developers and deployers should retain and make available to incident investigators. 

Effective incident analysis also requires robust institutional support.
Governments and regulators can establish secure reporting infrastructures, mandate the retention and sharing of key incident data, and build technical capacity for causal investigation--paralleling practices in aviation and finance. Regulatory frameworks such as the EU AI Act already include incident reporting requirements and can serve as a model for broader practices and mandates \citep{official_journal_of_the_european_union_regulation_2024}. By combining rigorous data collection and analysis with clear regulatory signals and public accountability mechanisms, we can strengthen our ability to learn from AI agent failures and prevent future harms. 

\section*{Acknowledgments}

The authors are grateful for discussions and feedback from the following individuals (in alphabetical order): Ben Bucknall, Stephen Casper, Jimmy Farrell, James Ginns, Joshua Greene, Noam Kolt, Shayne Longpre, Richard Moulange, Shalaleh Rismani, Jonas Schuett, Tommy Shaffer Shane, and Kevin Wei.

\appendix

\section{Taxonomy of Incident Factors}

  \subsection{System Factors}

\subsubsection{Training data}
\begin{itemize}
  \item Data poisoning
  \item False or misleading training data
    \item Misuse-enabling training data (e.g., dual-use content)
    \item Sensitive information in training data
  \item Content gaps in training (e.g., agent asked about a niche topic for which related training data was minimal)
    \end{itemize}
    \subsubsection{Feedback}
    \begin{itemize}
  \item Feedback specifies an undesirable proxy goal (e.g., maximizing human approval)
  \item Feedback improperly specifies appropriate trade-offs between goals (e.g., helpfulness and harmlessness)
  \item Poisoning feedback data
\end{itemize}

\subsubsection{Learning methods}
\begin{itemize}
  \item Undesirable memorization/knowledge retention (e.g., personally-identifiable information)
  \item Forgetting crucial information (i.e., `catastrophic forgetting'; \citealp{luo_empirical_2025})
  \item Draws undesirable inferences (e.g., links together personally identifiable information)
  \item Positive or negative reinforcement of the wrong goal/behavior (e.g., instrumental power-seeking, or `model collapse'; \citealp{shumailov_ai_2024})
  \item Does not robustly learn intended goal, capability/behavior, or knowledge (e.g., poor capability generalization)
\end{itemize}

\subsubsection{System prompt}
\begin{itemize}
  \item System prompt is underspecified
  \item System prompt inadvertently encourages undesirable behavior
\end{itemize}

\subsubsection{Scaffolding}
\begin{itemize}
  \item Scaffolding does not function as intended
  \item Scaffolding functions ineffectively (e.g., poor input sanitization)
\end{itemize}

      \subsection{Contextual Factors}
\subsubsection{Information}
\begin{itemize}
  \item Information is inaccessible to the AI agent (e.g., behind a paywall, or inaccessible format)
  \item Lack of existing information about a relevant topic
  \item Poor quality of information (e.g., false, outdated, or vague)
  \item Information environment is vulnerable to malicious inputs (e.g., prompt injections)
  \item Information environment includes sensitive data (e.g., personally-identifiable information)
\end{itemize}

\subsubsection{Tools}
\begin{itemize}
  \item Inability to access/use tools (e.g., incompatibilities, rate limits)
  \item Tools do not function as intended/anticipated
  \item Inability to conduct error monitoring of tools (e.g., view execution logs)
  \item Tools are vulnerable to attacks (e.g., prompt injections in tool instructions)
\end{itemize}

\subsubsection{Task definition}
\begin{itemize}
  \item Lack of task clarity (e.g., vague instructions)
  \item Task constraints (e.g., time limits, token limits, or other resource constraints)
  \item Conflicting task instructions (e.g., conflict between the system and user prompts)
\end{itemize}

    \subsection{Cognitive Errors}
\subsubsection{Observation}
\begin{itemize}
  \item Incorrect expectation of information
  \item Failure to notice/allocate attention to relevant information
  \item Unjustified allocation of attention to information (e.g., focusing on irrelevant information)
  \item Inadequate memory capacity (e.g., cannot process a large quantity of relevant information)
\end{itemize}

\subsubsection{Understanding}
\begin{itemize}
  \item Makes inferences/draws conclusions based on insufficient information
  \item Makes inferences/draws conclusions based on misleading information
  \item Makes inferences/draws conclusions based on incorrect assumptions
  \item Misinterpretation of information
\end{itemize}

\subsubsection{Decision-making}
\begin{itemize}
  \item Misinterpretation of user or developer intentions
  \item Incorrect selection and/or prioritization of goals (e.g., prioritizes user prompt over system prompt)
  \item Forgets or gets distracted from goal(s)
  \item Consideration of nonviable courses of action (e.g., makes a plan that depends upon access to a tool it cannot use)
  \item Insufficient consideration of alternative courses of action (e.g., improperly defaults to a heuristic)
  \item Improper evaluation criteria to choose between options (e.g., improper understanding of what actions are legal)
\end{itemize}

\subsubsection{Action execution}
\begin{itemize}
  \item Failure to take action (e.g., did not attempt)
  \item Failure of error monitoring during task performance (e.g., does not respond to feedback from tools)
  \item Incorrect execution of action (e.g., structures an API call improperly)
\end{itemize}

\newpage

\onecolumn

\section{Table of Open Questions and Challenges for Future Work}

\begin{table*}[ht!]
\centering
\label{tab:limitations}
\begin{tabular}{|p{2cm}|p{2.5cm}|p{9cm}|}
\hline
\textbf{Category} & \textbf{Limitation} & \textbf{Description} \\
\hline
\multirow{3}{2cm}{Framework Scope} 
& Non-exhaustive coverage & The proposed incident factors and information requirements are not exhaustive and may miss domain-specific factors \\
\cline{2-3}
& Emerging factors & Future trends could surface additional factors or render existing factors obsolete \\
\hline
\multirow{2}{2cm}{Theoretical Foundation} 
& High-level cognitive errors & The discussion of cognitive errors is necessarily high-level due to the nascent state of cognitive science of AI agents \\
\cline{2-3}
& Evolving understanding & Ongoing research into how AI agents interpret situations and make decisions may require changes to the framework \\
\hline
\multirow{4}{2cm}{Institutional Challenges} 
& Absent reporting infrastructure & No secure, standardized system currently exists for reporting AI incidents involving sensitive data \\
\cline{2-3}
& Expertise shortages & Organizations tasked with incident analysis (e.g. regulators) may lack sufficient AI technical expertise to investigate incident rigorously \\
\cline{2-3}
& Lack of accountability mechanisms & There are insufficient provisions to ensure that incident investigations remain accountable to the public--for example, through partial data release or periodic reports \\
\hline
\multirow{4}{2cm}{Privacy and Legal Constraints} 
& Sensitive data restrictions & Activity logs often contain personal or confidential information, limiting what details can be retained or shared publicly \\
\cline{2-3}
& Unclear legal landscape & Uncertain or unresolved legal issues (e.g. concerning liability or data protection) make it risky to store and share comprehensive incident data, discouraging full reporting \\
\cline{2-3}
& Limited data access & Privacy and security requirements restrict who can access sensitive incident information, narrowing the pool of people who can investigate and learn from incidents \\
\hline
\multirow{3}{2cm}{Operational Constraints} 
& Short data retention & Many AI developers have short or zero-retention policies, which can preclude collecting the information needed for comprehensive incident investigations \\
\cline{2-3}
& Reconstruction difficulties & Incident reconstruction may require access to proprietary system artifacts or model internals (white/grey-box access) that are not readily available to investigators \\
\cline{2-3}
& Voluntary adoption & Implementing the framework depends on voluntary cooperation by developers and deployers in the absence of regulations mandating compliance \\
\hline
\end{tabular}
\caption{Limitations and barriers to implementation of our framework}
\end{table*}

\twocolumn
\bibliography{references}

\begin{thebibliography}{75}
\providecommand{\natexlab}[1]{#1}

\bibitem[{{AIAAIC}(2025)}]{aiaaic_aiaaic_2025}
{AIAAIC}. 2025.
\newblock {AIAAIC}.
\newblock \url{https://www.aiaaic.org/}.
\newblock Accessed: 2025-05-21.

\bibitem[{{AIID}(2016)}]{aiid_incident_2016}
{AIID}. 2016.
\newblock Incident 106: {Korean} {Chatbot} {Luda} {Made} {Offensive} {Remarks} towards {Minority} {Groups}.
\newblock \url{https://incidentdatabase.ai/cite/106/}.
\newblock Accessed: 2025-05-24.

\bibitem[{{AIID}(2024)}]{aiid_welcome_2024}
{AIID}. 2024.
\newblock Welcome to the {Artificial} {Intelligence} {Incident} {Database}.
\newblock \url{https://incidentdatabase.ai/}.
\newblock Accessed: 2025-05-21.

\bibitem[{{Aim Labs Team}(2025)}]{aim_labs_team_breaking_2025}
{Aim Labs Team}. 2025.
\newblock Breaking down ‘{EchoLeak}’, the {First} {Zero}-{Click} {AI} {Vulnerability} {Enabling} {Data} {Exfiltration} from {Microsoft} 365 {Copilot}.
\newblock \url{https://www.aim.security/lp/aim-labs-echoleak-blogpost}.

\bibitem[{Alber et~al.(2025)Alber, Yang, Alyakin, Yang, Rai, Valliani, Zhang, Rosenbaum, Amend-Thomas, Kurland, Kremer, Eremiev, Negash, Wiggan, Nakatsuka, Sangwon, Neifert, Khan, Save, Palla, Grin, Hedman, Nasir-Moin, Liu, Jiang, Mankowski, Segev, Aphinyanaphongs, Riina, Golfinos, Orringer, Kondziolka, and Oermann}]{alber_medical_2025}
Alber, D.~A.; Yang, Z.; Alyakin, A.; Yang, E.; Rai, S.; Valliani, A.~A.; Zhang, J.; Rosenbaum, G.~R.; Amend-Thomas, A.~K.; Kurland, D.~B.; Kremer, C.~M.; Eremiev, A.; Negash, B.; Wiggan, D.~D.; Nakatsuka, M.~A.; Sangwon, K.~L.; Neifert, S.~N.; Khan, H.~A.; Save, A.~V.; Palla, A.; Grin, E.~A.; Hedman, M.; Nasir-Moin, M.; Liu, X.~C.; Jiang, L.~Y.; Mankowski, M.~A.; Segev, D.~L.; Aphinyanaphongs, Y.; Riina, H.~A.; Golfinos, J.~G.; Orringer, D.~A.; Kondziolka, D.; and Oermann, E.~K. 2025.
\newblock Medical large language models are vulnerable to data-poisoning attacks.
\newblock \emph{Nature Medicine}, 31(2): 618--626.
\newblock Publisher: Nature Publishing Group.

\bibitem[{{Anthropic}(2025{\natexlab{a}})}]{anthropic_how_2025}
{Anthropic}. 2025{\natexlab{a}}.
\newblock How long do you store my data?
\newblock \url{https://privacy.anthropic.com/en/articles/10023548-how-long-do-you-store-my-data}.
\newblock Accessed: 2025-05-21.

\bibitem[{{Anthropic}(2025{\natexlab{b}})}]{anthropic_how_2025-1}
{Anthropic}. 2025{\natexlab{b}}.
\newblock How long do you store my organization’s data?
\newblock \url{https://privacy.anthropic.com/en/articles/7996866-how-long-do-you-store-my-organization-s-data}.
\newblock Accessed: 2025-05-21.

\bibitem[{{Anthropic}(2025{\natexlab{c}})}]{anthropic_system-card_2025}
{Anthropic}. 2025{\natexlab{c}}.
\newblock System {Card}: {Claude} {Opus} 4 \& {Claude} {Sonnet} 4.
\newblock \url{https://www-cdn.anthropic.com/6be99a52cb68eb70eb9572b4cafad13df32ed995.pdf}.
\newblock Accessed: 2025-05-23.

\bibitem[{{Anthropic}(2025{\natexlab{d}})}]{anthropic_system_2025}
{Anthropic}. 2025{\natexlab{d}}.
\newblock System {Prompts}.
\newblock \url{https://docs.anthropic.com/en/release-notes/system-prompts}.
\newblock Accessed: 2025-05-21.

\bibitem[{{AVID}(2025)}]{avid_avid_2025}
{AVID}. 2025.
\newblock {AVID}.
\newblock \url{https://avidml.org/}.
\newblock Accessed: 2025-05-21.

\bibitem[{Bai et~al.(2022)Bai, Kadavath, Kundu, Askell, Kernion, Jones, Chen, Goldie, Mirhoseini, McKinnon, Chen, Olsson, Olah, Hernandez, Drain, Ganguli, Li, Tran-Johnson, Perez, Kerr, Mueller, Ladish, Landau, Ndousse, Lukosuite, Lovitt, Sellitto, Elhage, Schiefer, Mercado, DasSarma, Lasenby, Larson, Ringer, Johnston, Kravec, Showk, Fort, Lanham, Telleen-Lawton, Conerly, Henighan, Hume, Bowman, Hatfield-Dodds, Mann, Amodei, Joseph, McCandlish, Brown, and Kaplan}]{bai_constitutional_2022}
Bai, Y.; Kadavath, S.; Kundu, S.; Askell, A.; Kernion, J.; Jones, A.; Chen, A.; Goldie, A.; Mirhoseini, A.; McKinnon, C.; Chen, C.; Olsson, C.; Olah, C.; Hernandez, D.; Drain, D.; Ganguli, D.; Li, D.; Tran-Johnson, E.; Perez, E.; Kerr, J.; Mueller, J.; Ladish, J.; Landau, J.; Ndousse, K.; Lukosuite, K.; Lovitt, L.; Sellitto, M.; Elhage, N.; Schiefer, N.; Mercado, N.; DasSarma, N.; Lasenby, R.; Larson, R.; Ringer, S.; Johnston, S.; Kravec, S.; Showk, S.~E.; Fort, S.; Lanham, T.; Telleen-Lawton, T.; Conerly, T.; Henighan, T.; Hume, T.; Bowman, S.~R.; Hatfield-Dodds, Z.; Mann, B.; Amodei, D.; Joseph, N.; McCandlish, S.; Brown, T.; and Kaplan, J. 2022.
\newblock Constitutional {AI}: {Harmlessness} from {AI} {Feedback}.
\newblock arXiv:2212.08073.

\bibitem[{Baker et~al.(2025)Baker, Huizinga, Gao, Dou, Guan, Madry, Zaremba, Pachocki, and Farhi}]{baker_monitoring_2025}
Baker, B.; Huizinga, J.; Gao, L.; Dou, Z.; Guan, M.~Y.; Madry, A.; Zaremba, W.; Pachocki, J.; and Farhi, D. 2025.
\newblock Monitoring {Reasoning} {Models} for {Misbehavior} and the {Risks} of {Promoting} {Obfuscation}.
\newblock arXiv:2503.11926.

\bibitem[{Beurer-Kellner and Fischer(2025{\natexlab{a}})}]{beurer-kellner_introducing_2025}
Beurer-Kellner, L.; and Fischer, M. 2025{\natexlab{a}}.
\newblock Introducing {MCP}-{Scan}: {Protecting} {MCP} with {Invariant}.
\newblock \url{https://invariantlabs.ai/blog/introducing-mcp-scan}.

\bibitem[{Beurer-Kellner and Fischer(2025{\natexlab{b}})}]{beurer-kellner_mcp_2025}
Beurer-Kellner, L.; and Fischer, M. 2025{\natexlab{b}}.
\newblock {MCP} {Security} {Notification}: {Tool} {Poisoning} {Attacks}.
\newblock \url{https://invariantlabs.ai/blog/mcp-security-notification-tool-poisoning-attacks}.

\bibitem[{Biderman et~al.(2024)Biderman, Portes, Ortiz, Paul, Greengard, Jennings, King, Havens, Chiley, Frankle, Blakeney, and Cunningham}]{biderman_lora_2024}
Biderman, D.; Portes, J.; Ortiz, J. J.~G.; Paul, M.; Greengard, P.; Jennings, C.; King, D.; Havens, S.; Chiley, V.; Frankle, J.; Blakeney, C.; and Cunningham, J.~P. 2024.
\newblock {LoRA} {Learns} {Less} and {Forgets} {Less}.
\newblock \emph{Transactions on Machine Learning Research}.

\bibitem[{{Buck Shlegeris [@bshlgrs]}(2024)}]{buck_shlegeris_bshlgrs_i_2024}
{Buck Shlegeris [@bshlgrs]}. 2024.
\newblock I asked my {LLM} agent (a wrapper around {Claude} that lets it run bash commands and see their outputs): {\textgreater}can you ssh with the username buck to the computer on my network that is open to {SSH} because {I} didn’t know the local {IP} of my desktop. {I} walked away and promptly forgot {I}’d spun https://t.co/{I6qppMZFfk}.
\newblock \url{https://x.com/bshlgrs/status/1840577720465645960}.

\bibitem[{Carlini et~al.(2021)Carlini, Tramèr, Wallace, Jagielski, Herbert-Voss, Lee, Roberts, Brown, Song, Erlingsson, Oprea, and Raffel}]{carlini_extracting_2021}
Carlini, N.; Tramèr, F.; Wallace, E.; Jagielski, M.; Herbert-Voss, A.; Lee, K.; Roberts, A.; Brown, T.; Song, D.; Erlingsson, U.; Oprea, A.; and Raffel, C. 2021.
\newblock Extracting {Training} {Data} from {Large} {Language} {Models}.
\newblock 2633--2650.
\newblock ISBN 978-1-939133-24-3.

\bibitem[{Casper et~al.(2025)Casper, Bailey, Hunter, Ezell, Cabalé, Gerovitch, Slocum, Wei, Jurkovic, Khan, Christoffersen, Ozisik, Trivedi, Hadfield-Menell, and Kolt}]{casper_ai_2025}
Casper, S.; Bailey, L.; Hunter, R.; Ezell, C.; Cabalé, E.; Gerovitch, M.; Slocum, S.; Wei, K.; Jurkovic, N.; Khan, A.; Christoffersen, P. J.~K.; Ozisik, A.~P.; Trivedi, R.; Hadfield-Menell, D.; and Kolt, N. 2025.
\newblock The {AI} {Agent} {Index}.
\newblock arXiv:2502.01635.

\bibitem[{Casper et~al.(2023)Casper, Davies, Shi, Gilbert, Scheurer, Rando, Freedman, Korbak, Lindner, Freire, Wang, Marks, Segerie, Carroll, Peng, Christoffersen, Damani, Slocum, Anwar, Siththaranjan, Nadeau, Michaud, Pfau, Krasheninnikov, Chen, Langosco, Hase, Biyik, Dragan, Krueger, Sadigh, and Hadfield-Menell}]{casper_open_2023}
Casper, S.; Davies, X.; Shi, C.; Gilbert, T.~K.; Scheurer, J.; Rando, J.; Freedman, R.; Korbak, T.; Lindner, D.; Freire, P.; Wang, T.~T.; Marks, S.; Segerie, C.-R.; Carroll, M.; Peng, A.; Christoffersen, P. J.~K.; Damani, M.; Slocum, S.; Anwar, U.; Siththaranjan, A.; Nadeau, M.; Michaud, E.~J.; Pfau, J.; Krasheninnikov, D.; Chen, X.; Langosco, L.; Hase, P.; Biyik, E.; Dragan, A.; Krueger, D.; Sadigh, D.; and Hadfield-Menell, D. 2023.
\newblock Open {Problems} and {Fundamental} {Limitations} of {Reinforcement} {Learning} from {Human} {Feedback}.
\newblock \emph{Transactions on Machine Learning Research}.

\bibitem[{Casper et~al.(2024)Casper, Ezell, Siegmann, Kolt, Curtis, Bucknall, Haupt, Wei, Scheurer, Hobbhahn, Sharkey, Krishna, Von~Hagen, Alberti, Chan, Sun, Gerovitch, Bau, Tegmark, Krueger, and Hadfield-Menell}]{casper_black-box_2024}
Casper, S.; Ezell, C.; Siegmann, C.; Kolt, N.; Curtis, T.~L.; Bucknall, B.; Haupt, A.; Wei, K.; Scheurer, J.; Hobbhahn, M.; Sharkey, L.; Krishna, S.; Von~Hagen, M.; Alberti, S.; Chan, A.; Sun, Q.; Gerovitch, M.; Bau, D.; Tegmark, M.; Krueger, D.; and Hadfield-Menell, D. 2024.
\newblock Black-{Box} {Access} is {Insufficient} for {Rigorous} {AI} {Audits}.
\newblock In \emph{Proceedings of the 2024 {ACM} {Conference} on {Fairness}, {Accountability}, and {Transparency}}, {FAccT} '24, 2254--2272. New York, NY, USA: Association for Computing Machinery.
\newblock ISBN 9798400704505.

\bibitem[{Chan et~al.(2024)Chan, Ezell, Kaufmann, Wei, Hammond, Bradley, Bluemke, Rajkumar, Krueger, Kolt, Heim, and Anderljung}]{chan_visibility_2024}
Chan, A.; Ezell, C.; Kaufmann, M.; Wei, K.; Hammond, L.; Bradley, H.; Bluemke, E.; Rajkumar, N.; Krueger, D.; Kolt, N.; Heim, L.; and Anderljung, M. 2024.
\newblock Visibility into {AI} {Agents}.
\newblock In \emph{Proceedings of the 2024 {ACM} {Conference} on {Fairness}, {Accountability}, and {Transparency}}, {FAccT} '24, 958--973. New York, NY, USA: Association for Computing Machinery.
\newblock ISBN 9798400704505.

\bibitem[{Chan et~al.(2023)Chan, Salganik, Markelius, Pang, Rajkumar, Krasheninnikov, Langosco, He, Duan, Carroll, Lin, Mayhew, Collins, Molamohammadi, Burden, Zhao, Rismani, Voudouris, Bhatt, Weller, Krueger, and Maharaj}]{chan_harms_2023}
Chan, A.; Salganik, R.; Markelius, A.; Pang, C.; Rajkumar, N.; Krasheninnikov, D.; Langosco, L.; He, Z.; Duan, Y.; Carroll, M.; Lin, M.; Mayhew, A.; Collins, K.; Molamohammadi, M.; Burden, J.; Zhao, W.; Rismani, S.; Voudouris, K.; Bhatt, U.; Weller, A.; Krueger, D.; and Maharaj, T. 2023.
\newblock Harms from {Increasingly} {Agentic} {Algorithmic} {Systems}.
\newblock In \emph{Proceedings of the 2023 {ACM} {Conference} on {Fairness}, {Accountability}, and {Transparency}}, {FAccT} '23, 651--666. New York, NY, USA: Association for Computing Machinery.
\newblock ISBN 9798400701924.

\bibitem[{Creemers, Webster, and Toner(2022)}]{creemers_translation_2022}
Creemers, R.; Webster, G.; and Toner, H. 2022.
\newblock Translation: {Internet} {Information} {Service} {Algorithmic} {Recommendation} {Management} {Provisions} – {Effective} {March} 1, 2022.
\newblock \url{https://digichina.stanford.edu/work/translation-internet-information-service-algorithmic-recommendation-management-provisions-effective-march-1-2022/}.

\bibitem[{{European Commission}(2025)}]{european_commission_third_2025}
{European Commission}. 2025.
\newblock Third {Draft} of the {General}-{Purpose} {AI} {Code} of {Practice} published, written by independent experts.
\newblock \url{https://digital-strategy.ec.europa.eu/en/library/third-draft-general-purpose-ai-code-practice-published-written-independent-experts}.
\newblock Accessed: 2025-05-21.

\bibitem[{{European Union}(2024)}]{official_journal_of_the_european_union_regulation_2024}
{European Union}. 2024.
\newblock Regulation ({EU}) 2024/1689 of the {European} {Parliament} and of the {Council} of 13 {June} 2024 laying down harmonised rules on artificial intelligence and amending {Regulations} ({EC}) {No} 300/2008, ({EU}) {No} 167/2013, ({EU}) {No} 168/2013, ({EU}) 2018/858, ({EU}) 2018/1139 and ({EU}) 2019/2144 and {Directives} 2014/90/{EU}, ({EU}) 2016/797 and ({EU}) 2020/1828 ({Artificial} {Intelligence} {Act}) ({Text} with {EEA} relevance).
\newblock \url{http://data.europa.eu/eli/reg/2024/1689/oj/eng}.
\newblock Legislative Body: CONSIL, EP.

\bibitem[{Gabriel et~al.(2024)Gabriel, Manzini, Keeling, Hendricks, Rieser, Iqbal, Tomašev, Ktena, Kenton, Rodriguez, El-Sayed, Brown, Akbulut, Trask, Hughes, Bergman, Shelby, Marchal, Griffin, Mateos-Garcia, Weidinger, Street, Lange, Ingerman, Lentz, Enger, Barakat, Krakovna, Siy, Kurth-Nelson, McCroskery, Bolina, Law, Shanahan, Alberts, Balle, Haas, Ibitoye, Dafoe, Goldberg, Krier, Reese, Witherspoon, Hawkins, Rauh, Wallace, Franklin, Goldstein, Lehman, Klenk, Vallor, Biles, Morris, King, Arcas, Isaac, and Manyika}]{gabriel_ethics_2024}
Gabriel, I.; Manzini, A.; Keeling, G.; Hendricks, L.~A.; Rieser, V.; Iqbal, H.; Tomašev, N.; Ktena, I.; Kenton, Z.; Rodriguez, M.; El-Sayed, S.; Brown, S.; Akbulut, C.; Trask, A.; Hughes, E.; Bergman, A.~S.; Shelby, R.; Marchal, N.; Griffin, C.; Mateos-Garcia, J.; Weidinger, L.; Street, W.; Lange, B.; Ingerman, A.; Lentz, A.; Enger, R.; Barakat, A.; Krakovna, V.; Siy, J.~O.; Kurth-Nelson, Z.; McCroskery, A.; Bolina, V.; Law, H.; Shanahan, M.; Alberts, L.; Balle, B.; Haas, S.~d.; Ibitoye, Y.; Dafoe, A.; Goldberg, B.; Krier, S.; Reese, A.; Witherspoon, S.; Hawkins, W.; Rauh, M.; Wallace, D.; Franklin, M.; Goldstein, J.~A.; Lehman, J.; Klenk, M.; Vallor, S.; Biles, C.; Morris, M.~R.; King, H.; Arcas, B. A.~y.; Isaac, W.; and Manyika, J. 2024.
\newblock The {Ethics} of {Advanced} {AI} {Assistants}.
\newblock arXiv:2404.16244.

\bibitem[{Greshake et~al.(2023)Greshake, Abdelnabi, Mishra, Endres, Holz, and Fritz}]{greshake_not_2023}
Greshake, K.; Abdelnabi, S.; Mishra, S.; Endres, C.; Holz, T.; and Fritz, M. 2023.
\newblock Not {What} {You}'ve {Signed} {Up} {For}: {Compromising} {Real}-{World} {LLM}-{Integrated} {Applications} with {Indirect} {Prompt} {Injection}.
\newblock In \emph{Proceedings of the 16th {ACM} {Workshop} on {Artificial} {Intelligence} and {Security}}, {AISec} '23, 79--90. New York, NY, USA: Association for Computing Machinery.
\newblock ISBN 9798400702600.

\bibitem[{Grynbaum and Mac(2023)}]{grynbaum_times_2023}
Grynbaum, M.~M.; and Mac, R. 2023.
\newblock The {Times} {Sues} {OpenAI} and {Microsoft} {Over} {A}.{I}. {Use} of {Copyrighted} {Work}.
\newblock \emph{The New York Times}.

\bibitem[{Hagendorff et~al.(2024)Hagendorff, Dasgupta, Binz, Chan, Lampinen, Wang, Akata, and Schulz}]{hagendorff_machine_2024}
Hagendorff, T.; Dasgupta, I.; Binz, M.; Chan, S. C.~Y.; Lampinen, A.; Wang, J.~X.; Akata, Z.; and Schulz, E. 2024.
\newblock Machine {Psychology}.
\newblock arXiv:2303.13988.

\bibitem[{Hammond et~al.(2025)Hammond, Chan, Clifton, Hoelscher-Obermaier, Khan, McLean, Smith, Barfuss, Foerster, Gavenčiak, Han, Hughes, Kovařík, Kulveit, Leibo, Oesterheld, Witt, Shah, Wellman, Bova, Cimpeanu, Ezell, Feuillade-Montixi, Franklin, Kran, Krawczuk, Lamparth, Lauffer, Meinke, Motwani, Reuel, Conitzer, Dennis, Gabriel, Gleave, Hadfield, Haghtalab, Kasirzadeh, Krier, Larson, Lehman, Parkes, Piliouras, and Rahwan}]{hammond_multi-agent_2025}
Hammond, L.; Chan, A.; Clifton, J.; Hoelscher-Obermaier, J.; Khan, A.; McLean, E.; Smith, C.; Barfuss, W.; Foerster, J.; Gavenčiak, T.; Han, T.~A.; Hughes, E.; Kovařík, V.; Kulveit, J.; Leibo, J.~Z.; Oesterheld, C.; Witt, C. S.~d.; Shah, N.; Wellman, M.; Bova, P.; Cimpeanu, T.; Ezell, C.; Feuillade-Montixi, Q.; Franklin, M.; Kran, E.; Krawczuk, I.; Lamparth, M.; Lauffer, N.; Meinke, A.; Motwani, S.; Reuel, A.; Conitzer, V.; Dennis, M.; Gabriel, I.; Gleave, A.; Hadfield, G.; Haghtalab, N.; Kasirzadeh, A.; Krier, S.; Larson, K.; Lehman, J.; Parkes, D.~C.; Piliouras, G.; and Rahwan, I. 2025.
\newblock Multi-{Agent} {Risks} from {Advanced} {AI}.
\newblock arXiv:2502.14143.

\bibitem[{Hans et~al.(2024)Hans, Wen, Jain, Kirchenbauer, Kazemi, Singhania, Singh, Somepalli, Geiping, Bhatele, and Goldstein}]{hans_be_2024}
Hans, A.; Wen, Y.; Jain, N.; Kirchenbauer, J.; Kazemi, H.; Singhania, P.; Singh, S.; Somepalli, G.; Geiping, J.; Bhatele, A.; and Goldstein, T. 2024.
\newblock Be like a {Goldfish}, {Don}'t {Memorize}! {Mitigating} {Memorization} in {Generative} {LLMs}.
\newblock \emph{Advances in Neural Information Processing Systems}, 37: 24022--24045.

\bibitem[{Heiding et~al.(2024)Heiding, Lermen, Kao, Schneier, and Vishwanath}]{heiding_evaluating_2024}
Heiding, F.; Lermen, S.; Kao, A.; Schneier, B.; and Vishwanath, A. 2024.
\newblock Evaluating {Large} {Language} {Models}' {Capability} to {Launch} {Fully} {Automated} {Spear} {Phishing} {Campaigns}: {Validated} on {Human} {Subjects}.
\newblock arXiv:2412.00586.

\bibitem[{Hoffmann and Frase(2023)}]{hoffmann_adding_2023}
Hoffmann, M.; and Frase, H. 2023.
\newblock Adding {Structure} to {AI} {Harm}.
\newblock \url{https://cset.georgetown.edu/publication/adding-structure-to-ai-harm/}.
\newblock Accessed: 2025-05-21.

\bibitem[{Joshi et~al.(2025)Joshi, Shahid, Venneti, Vasu, Zheng, Li, Krishnamurthy, and Chan}]{joshi_coprompter_2025}
Joshi, I.; Shahid, S.; Venneti, S.~M.; Vasu, M.; Zheng, Y.; Li, Y.; Krishnamurthy, B.; and Chan, G. Y.-Y. 2025.
\newblock {CoPrompter}: {User}-{Centric} {Evaluation} of {LLM} {Instruction} {Alignment} for {Improved} {Prompt} {Engineering}.
\newblock In \emph{Proceedings of the 30th {International} {Conference} on {Intelligent} {User} {Interfaces}}, {IUI} '25, 341--365. New York, NY, USA: Association for Computing Machinery.
\newblock ISBN 9798400713064.

\bibitem[{Kapoor et~al.(2024)Kapoor, Stroebl, Siegel, Nadgir, and Narayanan}]{kapoor_ai_2024}
Kapoor, S.; Stroebl, B.; Siegel, Z.~S.; Nadgir, N.; and Narayanan, A. 2024.
\newblock {AI} {Agents} {That} {Matter}.
\newblock arXiv:2407.01502.

\bibitem[{Kasirzadeh and Gabriel(2025)}]{kasirzadeh_characterizing_2025}
Kasirzadeh, A.; and Gabriel, I. 2025.
\newblock Characterizing {AI} {Agents} for {Alignment} and {Governance}.
\newblock arXiv:2504.21848.

\bibitem[{Krakovna and Kramar(2023)}]{krakovna_power-seeking_2023}
Krakovna, V.; and Kramar, J. 2023.
\newblock Power-seeking can be probable and predictive for trained agents.
\newblock arXiv:2304.06528.

\bibitem[{Krueger, Maharaj, and Leike(2020)}]{krueger_hidden_2020}
Krueger, D.; Maharaj, T.; and Leike, J. 2020.
\newblock Hidden {Incentives} for {Auto}-{Induced} {Distributional} {Shift}.
\newblock arXiv:2009.09153.

\bibitem[{Lai et~al.(2023)Lai, Ngo, Pouran Ben~Veyseh, Man, Dernoncourt, Bui, and Nguyen}]{lai_chatgpt_2023}
Lai, V.~D.; Ngo, N.; Pouran Ben~Veyseh, A.; Man, H.; Dernoncourt, F.; Bui, T.; and Nguyen, T.~H. 2023.
\newblock {ChatGPT} {Beyond} {English}: {Towards} a {Comprehensive} {Evaluation} of {Large} {Language} {Models} in {Multilingual} {Learning}.
\newblock In Bouamor, H.; Pino, J.; and Bali, K., eds., \emph{Findings of the {Association} for {Computational} {Linguistics}: {EMNLP} 2023}, 13171--13189. Singapore: Association for Computational Linguistics.

\bibitem[{Leveson(2023)}]{leveson_introduction_2023}
Leveson, N.~G. 2023.
\newblock \emph{An {Introduction} to {System} {Safety} {Engineering}}.
\newblock Cambridge, Massachusetts: The MIT Press.
\newblock ISBN 978-0-262-54688-1.

\bibitem[{Lin, Hilton, and Evans(2022)}]{lin_truthfulqa_2022}
Lin, S.; Hilton, J.; and Evans, O. 2022.
\newblock {TruthfulQA}: {Measuring} {How} {Models} {Mimic} {Human} {Falsehoods}.
\newblock In Muresan, S.; Nakov, P.; and Villavicencio, A., eds., \emph{Proceedings of the 60th {Annual} {Meeting} of the {Association} for {Computational} {Linguistics} ({Volume} 1: {Long} {Papers})}, 3214--3252. Dublin, Ireland: Association for Computational Linguistics.

\bibitem[{Longpre et~al.(2025)Longpre, Klyman, Appel, Kapoor, Bommasani, Sahar, McGregor, Ghosh, Blili-Hamelin, Butters, Nelson, Elazari, Sellars, Ellis, Sherrets, Song, Geiger, Cohen, McIlvenny, Srikumar, Jaycox, Anderljung, Johnson, Carlini, Miailhe, Marda, Henderson, Portnoff, Weiss, Westerhoff, Jernite, Chowdhury, Liang, and Narayanan}]{longpre_-house_2025}
Longpre, S.; Klyman, K.; Appel, R.~E.; Kapoor, S.; Bommasani, R.; Sahar, M.; McGregor, S.; Ghosh, A.; Blili-Hamelin, B.; Butters, N.; Nelson, A.; Elazari, A.; Sellars, A.; Ellis, C.~J.; Sherrets, D.; Song, D.; Geiger, H.; Cohen, I.; McIlvenny, L.; Srikumar, M.; Jaycox, M.~M.; Anderljung, M.; Johnson, N.~F.; Carlini, N.; Miailhe, N.; Marda, N.; Henderson, P.; Portnoff, R.~S.; Weiss, R.; Westerhoff, V.; Jernite, Y.; Chowdhury, R.; Liang, P.; and Narayanan, A. 2025.
\newblock In-{House} {Evaluation} {Is} {Not} {Enough}: {Towards} {Robust} {Third}-{Party} {Flaw} {Disclosure} for {General}-{Purpose} {AI}.
\newblock arXiv:2503.16861.

\bibitem[{Luo et~al.(2025)Luo, Yang, Meng, Li, Zhou, and Zhang}]{luo_empirical_2025}
Luo, Y.; Yang, Z.; Meng, F.; Li, Y.; Zhou, J.; and Zhang, Y. 2025.
\newblock An {Empirical} {Study} of {Catastrophic} {Forgetting} in {Large} {Language} {Models} {During} {Continual} {Fine}-tuning.
\newblock arXiv:2308.08747.

\bibitem[{McCoy et~al.(2024)McCoy, Yao, Friedman, Hardy, and Griffiths}]{mccoy_embers_2024}
McCoy, R.~T.; Yao, S.; Friedman, D.; Hardy, M.~D.; and Griffiths, T.~L. 2024.
\newblock Embers of autoregression show how large language models are shaped by the problem they are trained to solve.
\newblock \emph{Proceedings of the National Academy of Sciences}, 121(41): e2322420121.
\newblock Publisher: Proceedings of the National Academy of Sciences.

\bibitem[{Meng et~al.(2022)Meng, Bau, Andonian, and Belinkov}]{meng_locating_2022}
Meng, K.; Bau, D.; Andonian, A.; and Belinkov, Y. 2022.
\newblock Locating and editing factual associations in {GPT}.
\newblock In \emph{Proceedings of the 36th {International} {Conference} on {Neural} {Information} {Processing} {Systems}}, {NIPS} '22, 17359--17372. Red Hook, NY, USA: Curran Associates Inc.
\newblock ISBN 978-1-71387-108-8.

\bibitem[{{Microsoft}(2025)}]{microsoft_m365_2025}
{Microsoft}. 2025.
\newblock M365 {Copilot} {Information} {Disclosure} {Vulnerability}.
\newblock \url{https://msrc.microsoft.com/update-guide/en-US/vulnerability/CVE-2025-32711}.
\newblock Accessed: 2025-08-04.

\bibitem[{Mitchell et~al.(2019)Mitchell, Wu, Zaldivar, Barnes, Vasserman, Hutchinson, Spitzer, Raji, and Gebru}]{mitchell_model_2019}
Mitchell, M.; Wu, S.; Zaldivar, A.; Barnes, P.; Vasserman, L.; Hutchinson, B.; Spitzer, E.; Raji, I.~D.; and Gebru, T. 2019.
\newblock Model {Cards} for {Model} {Reporting}.
\newblock In \emph{Proceedings of the {Conference} on {Fairness}, {Accountability}, and {Transparency}}, {FAT}* '19, 220--229. New York, NY, USA: Association for Computing Machinery.
\newblock ISBN 978-1-4503-6125-5.

\bibitem[{{Occupational Safety and Health Administration}(n.d.)}]{occupational_safety_and_health_administration_incident_nodate}
{Occupational Safety and Health Administration}. n.d.
\newblock Incident {Investigation}.
\newblock \url{https://www.osha.gov/incident-investigation}.
\newblock Accessed: 2025-08-04.

\bibitem[{{OECD}(2025{\natexlab{a}})}]{oecd_oecd_2025}
{OECD}. 2025{\natexlab{a}}.
\newblock {OECD} {AI} {Incidents} {Monitor}, an evidence base for trustworthy {AI}.
\newblock \url{https://oecd.ai/en/incidents}.
\newblock Accessed: 2025-05-21.

\bibitem[{{OECD}(2025{\natexlab{b}})}]{oecd_towards_2025}
{OECD}. 2025{\natexlab{b}}.
\newblock Towards a common reporting framework for {AI} incidents.
\newblock Technical Report~34, OECD Publishing, Paris.

\bibitem[{O'Keefe et~al.(2025)O'Keefe, Ramakrishnan, Tay, and Winter}]{okeefe_law-following_2025}
O'Keefe, C.; Ramakrishnan, K.; Tay, J.; and Winter, C. 2025.
\newblock Law-{Following} {AI}: {Designing} {AI} {Agents} to {Obey} {Human} {Laws}.
\newblock SSRN:5242643.

\bibitem[{{OpenAI}(2024{\natexlab{a}})}]{openai_enterprise_2024}
{OpenAI}. 2024{\natexlab{a}}.
\newblock Enterprise privacy at {OpenAI}.
\newblock \url{https://openai.com/enterprise-privacy/}.
\newblock Accessed: 2025-05-21.

\bibitem[{{OpenAI}(2024{\natexlab{b}})}]{openai_openai_2024}
{OpenAI}. 2024{\natexlab{b}}.
\newblock {OpenAI} o1 {System} {Card}.
\newblock \url{https://cdn.openai.com/o1-system-card-20241205.pdf}.
\newblock Accessed: 2025-05-23.

\bibitem[{{OpenAI}(2025{\natexlab{a}})}]{openai_data_2025}
{OpenAI}. 2025{\natexlab{a}}.
\newblock Data controls in the {OpenAI} platform.
\newblock \url{https://platform.openai.com}.
\newblock Accessed: 2025-05-23.

\bibitem[{{OpenAI}(2025{\natexlab{b}})}]{openai_expanding_2025}
{OpenAI}. 2025{\natexlab{b}}.
\newblock Expanding on what we missed with sycophancy.
\newblock \url{https://openai.com/index/expanding-on-sycophancy/}.

\bibitem[{{OpenAI}(2025{\natexlab{c}})}]{openai_model_2025}
{OpenAI}. 2025{\natexlab{c}}.
\newblock Model behavior feedback.
\newblock \url{https://openai.com/form/model-behavior-feedback/}.
\newblock Accessed: 2025-05-24.

\bibitem[{{OpenAI}(2025{\natexlab{d}})}]{openai_sycophancy_2025}
{OpenAI}. 2025{\natexlab{d}}.
\newblock Sycophancy in {GPT}-4o: {What} happened and what we’re doing about it.
\newblock \url{https://openai.com/index/sycophancy-in-gpt-4o/}.

\bibitem[{Ouyang et~al.(2022)Ouyang, Wu, Jiang, Almeida, Wainwright, Mishkin, Zhang, Agarwal, Slama, Ray, Schulman, Hilton, Kelton, Miller, Simens, Askell, Welinder, Christiano, Leike, and Lowe}]{ouyang_training_2022}
Ouyang, L.; Wu, J.; Jiang, X.; Almeida, D.; Wainwright, C.~L.; Mishkin, P.; Zhang, C.; Agarwal, S.; Slama, K.; Ray, A.; Schulman, J.; Hilton, J.; Kelton, F.; Miller, L.; Simens, M.; Askell, A.; Welinder, P.; Christiano, P.; Leike, J.; and Lowe, R. 2022.
\newblock Training language models to follow instructions with human feedback.
\newblock In \emph{Proceedings of the 36th {International} {Conference} on {Neural} {Information} {Processing} {Systems}}, {NIPS} '22, 27730--27744. Red Hook, NY, USA: Curran Associates Inc.
\newblock ISBN 978-1-71387-108-8.

\bibitem[{Paeth et~al.(2025)Paeth, Atherton, Pittaras, Frase, and McGregor}]{paeth_lessons_2025}
Paeth, K.; Atherton, D.; Pittaras, N.; Frase, H.; and McGregor, S. 2025.
\newblock Lessons for {Editors} of {AI} {Incidents} from the {AI} {Incident} {Database}.
\newblock \emph{Proceedings of the AAAI Conference on Artificial Intelligence}, 39(28): 28946--28953.
\newblock Number: 28.

\bibitem[{Pittaras and McGregor(2022)}]{pittaras_taxonomic_2022}
Pittaras, N.; and McGregor, S. 2022.
\newblock A taxonomic system for failure cause analysis of open source {AI} incidents.
\newblock arXiv:2211.07280.

\bibitem[{Preston(2025)}]{preston_groks_2025}
Preston, D. 2025.
\newblock Grok’s white genocide fixation caused by ‘unauthorized modification’.
\newblock \url{https://www.theverge.com/news/668220/grok-white-genocide-south-africa-xai-unauthorized-modification-employee}.

\bibitem[{Ranaldi and Pucci(2024)}]{ranaldi_when_2024}
Ranaldi, L.; and Pucci, G. 2024.
\newblock When {Large} {Language} {Models} contradict humans? {Large} {Language} {Models}' {Sycophantic} {Behaviour}.
\newblock arXiv:2311.09410.

\bibitem[{Shappell and Wiegmann(2000)}]{shappell_human_2000}
Shappell, S.~A.; and Wiegmann, D.~A. 2000.
\newblock Human {Factors} {Analysis} and {Classification} {System}-{HFACS}.
\newblock Technical Report DOT/FAA/AM-00/7.

\bibitem[{Sharma et~al.(2023)Sharma, Tong, Korbak, Duvenaud, Askell, Bowman, Durmus, Hatfield-Dodds, Johnston, Kravec, Maxwell, McCandlish, Ndousse, Rausch, Schiefer, Yan, Zhang, and Perez}]{sharma_towards_2023}
Sharma, M.; Tong, M.; Korbak, T.; Duvenaud, D.; Askell, A.; Bowman, S.~R.; Durmus, E.; Hatfield-Dodds, Z.; Johnston, S.~R.; Kravec, S.~M.; Maxwell, T.; McCandlish, S.; Ndousse, K.; Rausch, O.; Schiefer, N.; Yan, D.; Zhang, M.; and Perez, E. 2023.
\newblock Towards {Understanding} {Sycophancy} in {Language} {Models}.

\bibitem[{Shumailov et~al.(2024)Shumailov, Shumaylov, Zhao, Papernot, Anderson, and Gal}]{shumailov_ai_2024}
Shumailov, I.; Shumaylov, Z.; Zhao, Y.; Papernot, N.; Anderson, R.; and Gal, Y. 2024.
\newblock {AI} models collapse when trained on recursively generated data.
\newblock \emph{Nature}, 631(8022): 755--759.
\newblock Publisher: Nature Publishing Group.

\bibitem[{Tlaie and Farrell(2025)}]{tlaie_securing_2025}
Tlaie, A.; and Farrell, J. 2025.
\newblock Securing {External} {Deeper}-than-black-box {GPAI} {Evaluations}.
\newblock arXiv:2503.07496.

\bibitem[{Turpin et~al.(2023)Turpin, Michael, Perez, and Bowman}]{turpin_language_2023}
Turpin, M.; Michael, J.; Perez, E.; and Bowman, S. 2023.
\newblock Language {Models} {Don}'t {Always} {Say} {What} {They} {Think}: {Unfaithful} {Explanations} in {Chain}-of-{Thought} {Prompting}.
\newblock \emph{Advances in Neural Information Processing Systems}, 36: 74952--74965.

\bibitem[{{UK AI Security Institute}(2024)}]{uk_ai_security_institute_pre-deployment_2024}
{UK AI Security Institute}. 2024.
\newblock Pre-{Deployment} {Evaluation} of {OpenAI}’s o1 {Model}.
\newblock \url{https://www.aisi.gov.uk/work/pre-deployment-evaluation-of-openais-o1-model}.

\bibitem[{Villalobos and Atkinson(2023)}]{villalobos_trading_2023}
Villalobos, P.; and Atkinson, D. 2023.
\newblock Trading {Off} {Compute} in {Training} and {Inference}.
\newblock \url{https://epoch.ai/blog/trading-off-compute-in-training-and-inference}.

\bibitem[{Wallace et~al.(2024)Wallace, Xiao, Leike, Weng, Heidecke, and Beutel}]{wallace_instruction_2024}
Wallace, E.; Xiao, K.; Leike, R.; Weng, L.; Heidecke, J.; and Beutel, A. 2024.
\newblock The {Instruction} {Hierarchy}: {Training} {LLMs} to {Prioritize} {Privileged} {Instructions}.
\newblock arXiv:2404.13208.

\bibitem[{Wei, Haghtalab, and Steinhardt(2023)}]{wei_jailbroken_2023}
Wei, A.; Haghtalab, N.; and Steinhardt, J. 2023.
\newblock Jailbroken: {How} {Does} {LLM} {Safety} {Training} {Fail}?

\bibitem[{Weidinger et~al.(2022)Weidinger, Uesato, Rauh, Griffin, Huang, Mellor, Glaese, Cheng, Balle, Kasirzadeh, Biles, Brown, Kenton, Hawkins, Stepleton, Birhane, Hendricks, Rimell, Isaac, Haas, Legassick, Irving, and Gabriel}]{weidinger_taxonomy_2022}
Weidinger, L.; Uesato, J.; Rauh, M.; Griffin, C.; Huang, P.-S.; Mellor, J.; Glaese, A.; Cheng, M.; Balle, B.; Kasirzadeh, A.; Biles, C.; Brown, S.; Kenton, Z.; Hawkins, W.; Stepleton, T.; Birhane, A.; Hendricks, L.~A.; Rimell, L.; Isaac, W.; Haas, J.; Legassick, S.; Irving, G.; and Gabriel, I. 2022.
\newblock Taxonomy of {Risks} posed by {Language} {Models}.
\newblock In \emph{Proceedings of the 2022 {ACM} {Conference} on {Fairness}, {Accountability}, and {Transparency}}, {FAccT} '22, 214--229. New York, NY, USA: Association for Computing Machinery.
\newblock ISBN 978-1-4503-9352-2.

\bibitem[{Welleck et~al.(2024)Welleck, Bertsch, Finlayson, Schoelkopf, Xie, Neubig, Kulikov, and Harchaoui}]{welleck_decoding_2024}
Welleck, S.; Bertsch, A.; Finlayson, M.; Schoelkopf, H.; Xie, A.; Neubig, G.; Kulikov, I.; and Harchaoui, Z. 2024.
\newblock From {Decoding} to {Meta}-{Generation}: {Inference}-time {Algorithms} for {Large} {Language} {Models}.
\newblock \emph{Transactions on Machine Learning Research}.

\bibitem[{Xing et~al.(2017)Xing, Parry, Presley, Forester, Hendrickson, and Dang}]{xing_integrated_2017}
Xing, J.; Parry, G.; Presley, M.; Forester, J.; Hendrickson, S.; and Dang, V. 2017.
\newblock An {Integrated} {Human} {Event} {Analysis} {System} ({IDHEAS}) for {Nuclear} {Power} {Plant} {Internal} {Events} {At}-{Power}.
\newblock Technical Report NUREG-2199, U.S. Nuclear Regulatory Commission, Washington, D.C.

\bibitem[{Yang et~al.(2024)Yang, Sun, Ma, Liu, Yin, and Cheng}]{yang_butterfly_2024}
Yang, W.; Sun, F.; Ma, X.; Liu, X.; Yin, D.; and Cheng, X. 2024.
\newblock The {Butterfly} {Effect} of {Model} {Editing}: {Few} {Edits} {Can} {Trigger} {Large} {Language} {Models} {Collapse}.
\newblock In Ku, L.-W.; Martins, A.; and Srikumar, V., eds., \emph{Findings of the {Association} for {Computational} {Linguistics}: {ACL} 2024}, 5419--5437. Bangkok, Thailand: Association for Computational Linguistics.

\end{thebibliography}
\end{document}